# Recent Progress in Studies of Cobalt-based Quasi-1-dimensional Quantum Magnets

Lun Jin[1, 2, *] and Robert J. Cava[2, *]

[1]Key Laboratory of Quantum Materials and Devices of Ministry of Education, School of Physics, Southeast University, Nanjing 211189, China

[2]Department of Chemistry, Princeton University, Princeton, New Jersey 08544, USA

* E-mails of corresponding authors: ljin@seu.edu.cn; rcava@princeton.edu

**Abstract:** The interplay of crystal electric field, temperature, and spin-orbit coupling can yield a Kramer's ion and thus an effective $S = ½$ ground state for $Co^{2+}$ ions ($3d^7$), which is often the case for low dimensional materials. This is because a highly anisotropic structural motif can force the spins to point either "up" or "down", hence becoming a system where spins communicate via Ising interactions. Cobalt-based quasi-1-dimensional materials have been studied in this context since the latter half of the 20$^{th}$ century, but due to the development of modern characterization techniques and advances in sample preparation, the exotic physical phenomena that have generated the most interest have only emerged in the most recent three to four decades. This topical review mainly summarizes progress in cobalt-based quasi-1-dimensional quantum magnets, and comments on a few research directions of potential future interest.

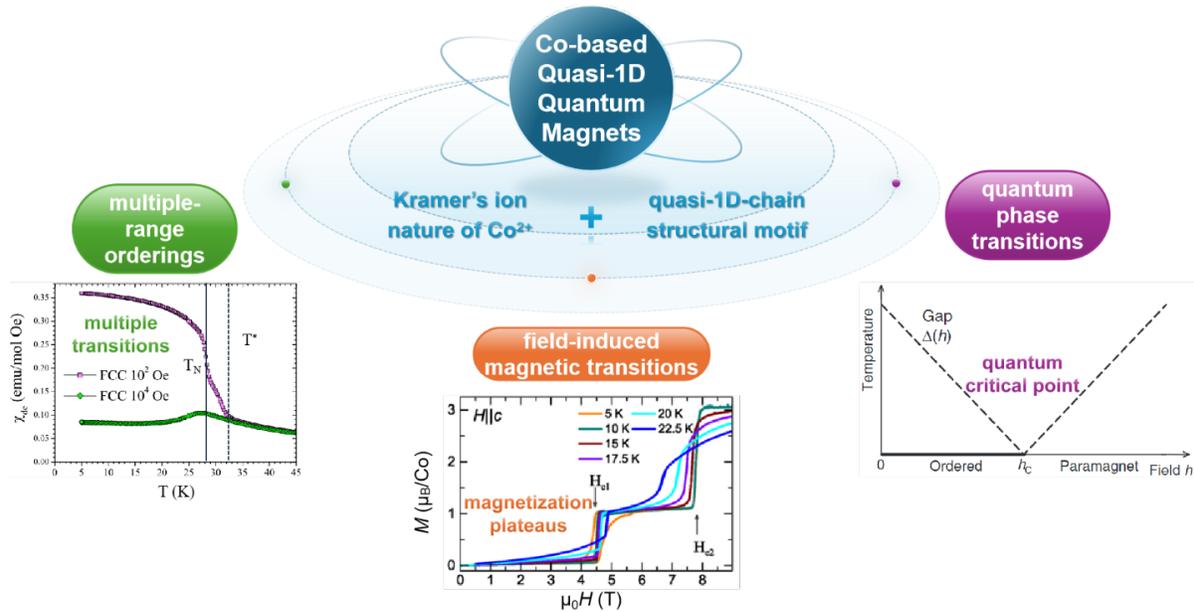

# 1. Introduction

Cobalt-based quasi-1-dimensional materials show complex magnetism in the quantum regime, and thus have attracted enduring interest in materials physics. Their properties can emerge from the interplay of the Kramer's ion nature of $Co^{2+}$ ions ($3d^7$) and the quasi-1D-chain structural motif of the materials. To best understand the exotic phenomena they display, one must start from the basics. The magnetic $Co^{2+}$ ion ($3d^7$) in the lattice is forced to point either "up" or "down" (effective spin-½) along the easy axis by a strong single-ion anisotropy, hence it can become a two-level system that allows spins to communicate via Ising interactions. [1,2] $CoO_6$ octahedra are connected by edge-sharing or face-sharing within the 1D chains, resulting in shorter distances and smaller Co–O–Co bond angles between adjacent Co centers compared to the more commonly seen corner-sharing linkage. In some cases, these 1D chains also arrange themselves into two-dimensional (2D) lattices (e.g., the triangular lattice [3].) that can enable geometric frustration. Therefore, the dimensionality of this class of materials (i.e., the relative strengths of intra-chain and inter-chain magnetic couplings) can be tuned over a wide range, owing to the fruitful choices in the way of bridging these Co-containing 1D chains together.

The competition between intra-chain and inter-chain magnetic couplings can lead to complicated ordered and/or disordered states in cobalt-based quasi-1D quantum magnets. For example, multiple magnetic transitions can sometimes be observed in temperature-dependent magnetization data [see e.g. 4,5], and plateaus can be observed in field-dependent magnetization data [see e.g. 6,7]. The superposition of short- to medium-range ordering in addition to long-range ordering [5,8], and the transition from commensurate to incommensurate phases [9,10] also emerges from time to time. Besides these phenomena, one of the most astonishing features displayed is a quantum phase transition (QPT) in some of the materials discussed below, just as was theoretically predicted decades ago [11]. Unlike a classical phase transition driven by thermal fluctuations, quantum phase transitions are defined as a qualitative change in the ground state of a system that occurs upon modifying a non-thermal external parameter (such as magnetic field, pressure, or doping) at a temperature of zero Kelvin. [12] The 1D quantum Ising chain in a transverse magnetic field is one of the most studied examples of a continuous quantum phase transition; this is because the zero-Kelvin ground state can be extrapolated to finite temperatures, hence making it viable to test fundamental ideas in real materials and study emergent physical properties. [13,14]

Although the study of the quantum Ising chain was addressed more than 50 years ago [15], cobalt-based quasi-1D quantum magnets exhibiting rich physics have not been fully understood even up to date and remain one of the keen research topics in the field. This topical review aims to summarize attempts that were made by both experimentalists and theorists along the journey, and to describe and rationalize spectacular findings in various types of cobalt-based quasi-1D quantum magnets. Hopefully by facilitating cross-references between different materials in this category, new leads can emerge and therefore encourage further investigations regarding this research topic.

## 2. One-dimensional Chains Consisting of Edge-sharing $CoO_6$ Polyhedra

### 2.1 Co-containing Columbites

The most famous member of the columbite family is probably $CoNb_2O_6$, which has been extensively studied for decades within the physics community. Its crystal structure consists of zig-zag chains running along the $c$ axis that are made of edge-sharing oxygen octahedra containing either Co or Nb. Early studies mainly focused on the complicated ordered magnetic states of $CoNb_2O_6$ in the low temperature regime, using techniques including neutron scattering, specific heat capacity, magnetic susceptibility and magnetization. It is reported that the material has two intrinsic magnetic transitions, at $T_1 \sim 3.0$ K and $T_2 \sim 1.9$ K, with the former one associated with a 3D helical magnetic ordered phase while the latter one exhibits a pronounced 2D character even in the 3D ordered state. [4,16–18]. The ferromagnetic Ising chains, containing $Co^{2+}$, are only weakly coupled to each other as they are separated by two rows of non-magnetic $NbO_6$ chains (**Fig. 1a**). (If $J_0$ denotes the intra-chain ferromagnetic exchange along the easy axis, which is in the $a$–$c$ plane at an angle $\pm 31°$ to the $c$ axis (**Fig. 1b**) and $J_1$ and $J_2$ denote the much weaker inter-chain antiferromagnetic interactions then $|J_1|, |J_2| \ll J_0$) (**Fig. 1c**)). [2,19] Researchers have worked on the $H_{//c}$–$T$ magnetic phase diagram of $CoNb_2O_6$, revealing antiferromagnetic, ferrimagnetic, and incommensurate phases, plus various ordered phases that are field-induced between the ferrimagnetic and saturated paramagnetic phases. [9]

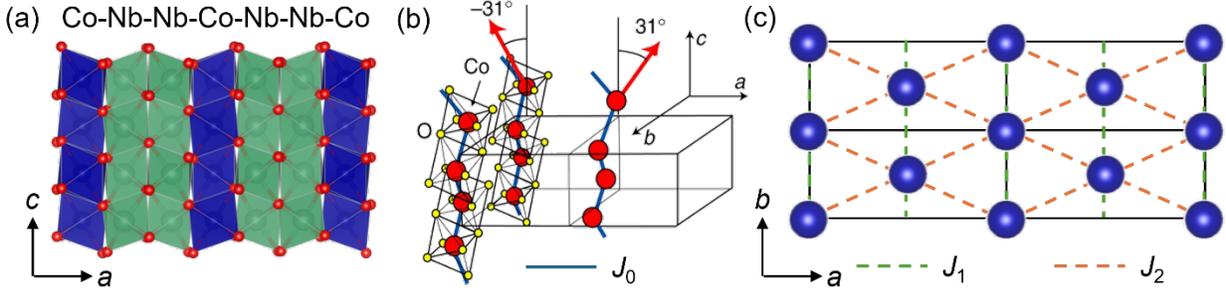

**Figure 1. Crystal and magnetic structure of $CoNb_2O_6$ columbite.** (a) the unit cell; (b) the zig-zag 1D chain made of edge-sharing $CoO_6$ octahedra (red arrows indicate the easy axis) [2] with $J_0$ denotes the intra-chain ferromagnetic exchange; (c) the isosceles-triangular-lattice in the *a-b* plane with $J_1$ and $J_2$ denotes the much weaker inter-chain antiferromagnetic interactions.

$CoNb_2O_6$ is considered as an ideal system to study the interplay of quantum fluctuations and quantum criticality, which are two of the most intriguing topics in current condensed-matter physics. As illustrated above, the Ising chains in $CoNb_2O_6$ are arranged in a triangular lattice, which is archetypal for geometrically frustrated magnets, and hence quantum fluctuations can be induced in the system. [1] Thus $^{93}$Nb nuclear magnetic resonance (NMR) was used to investigate the quantum spin fluctuations in $CoNb_2O_6$ near the quantum critical point (QCP), owing to the hyperfine interactions between Co electron spins and $^{93}$Nb nuclear spins. [20] In addition, $CoNb_2O_6$ has been found to be a close realization of the transverse Ising magnet (TIM) in a real material. In the absence of a transverse field, the weakly coupled ferromagnetic Ising chains remain in ordered states (spins aligned "up" or "down" along the Ising axis), but when a transverse field that can overcome the exchange interactions is applied, the system undergoes a phase transition to a disordered state at a QCP, attributed to the Ising spins all lying in the quantum superposition of "up" and "down" states. [1,2] Heat capacity results show that the spin entropy is largely enhanced at the QCP and find evidence for gapless spin excitations that are both participating in, and affected by, the quantum phase transition. [2] Anomalous lattice softening is also observed near the QCP in $CoNb_2O_6$, due to the relativistic spin-orbit interaction in the quantum critical regime.[21] Inelastic neutron scattering experiments have provided direct evidence for quantum criticality in the Ising chains by using strong transverse fields to tune $CoNb_2O_6$ through its QCP. [14] In complement to the five kink bound states observed by neutron scattering [14], high-resolution time-domain terahertz (THz) spectroscopy finds more features in the far infrared for $CoNb_2O_6$,

with four additional kink bound states and a new higher energy excitation below the continuum resolved. [22]

However, with more experimental studies about the ordered/disordered states of $CoNb_2O_6$ coming out, substantial deviations from the ideal transverse-field Ising model were noted. For example, domain walls in $CoNb_2O_6$ display quantum motion at zero applied field, which is inconsistent with a pure 1D Ising model. [9,14,17,23,24] This feature was probed by a theoretical approach, using a microscopic spin-exchange Hamiltonian to demonstrate the crucial role of glide symmetry breaking in the system. [25] Several experimental studies have also focused on the domain-wall confinement/freezing in $CoNb_2O_6$. [26–28] A 'twisted Kitaev chain' composed of a bond-dependent Ising interaction was proposed for $CoNb_2O_6$, inspired by similar interactions found in those honeycomb Kitaev spin liquids. [29] This idea was recently backed up by theorists based on the *JKΓ* model (Heisenberg (*J*)-Kitaev (*K*)-Gamma (*Γ*)) for a $3d^7$ systems such as $Co^{2+}$. They mapped out a microscopic origin for Ising behavior in spin-orbit coupled 1D chains and posit $CoNb_2O_6$ as a rare Kitaev chain. [30]

$CoTa_2O_6$ (**Fig. 2a**) and $CoV_2O_6$ (**Fig. 2b-c**) have also been studied, but not to a comparable extent so far. There are very limited papers on $CoTa_2O_6$ [31–33], but one study did grow a single crystal of this compound by using the optical floating-zone technique [34]. Its cation-ordered rutile-derived structure results in a Co-Ta-Ta-Co sequence within each 1D chain, which might dilute the intra-chain Co-Co magnetic coupling to a certain extent and forfeit some exotic phenomena. Generally speaking, $CoV_2O_6$ has received more attention from the community than $CoTa_2O_6$. It has two structural forms – a low-temperature phase (triclinic γ-$CoV_2O_6$, **Fig. 2b**) and a high-temperature phase (monoclinic α-$CoV_2O_6$, **Fig. 2c**), separated by a structural transition which occurs at 953 K. [35] Upon resolving the magnetic structure from neutron powder diffraction data, it was found that the γ phase contains two independent Co sites which are not perfectly aligned but show angles different from 180°, and thus cannot unambiguously show ferromagnetic order inside the chains. [36] In contrast, the magnetic moments in the α phase lie in the *ac* plane and are coupled within the chains running along the *b* axis, no matter in the ground or field-induced states. [37–40] Attempts to grow $CoV_2O_6$ single crystals have been made by using both the flux [35,41] and the floating-zone techniques. [42] Very recently, however, quantum annealing resulting from time-reversal symmetry breaking in a tiny transverse field was reported in a single-crystal study of α-$CoV_2O_6$. The many-body simulations in that work point out that a tiny applied transverse field can profoundly enhance quantum spin fluctuations within the system. [43] Thus, α-CoV$_2$O$_6$ could be analogous to its sister compound CoNb$_2$O$_6$ in a certain way, with a potential superiority – the easement in the strength of the applied transverse field. The above-mentioned results in the literature suggest that exotic physical phenomena can be seen in CoV$_2$O$_6$ as well, as long as high-quality samples can be synthesized. In other words, obstacles such as the competition between polymorphs or the volatility of starting reagents during the crystal growth process need to be overcome, hence yielding the much needed "ultra-clean" samples for further study in the quantum regime.

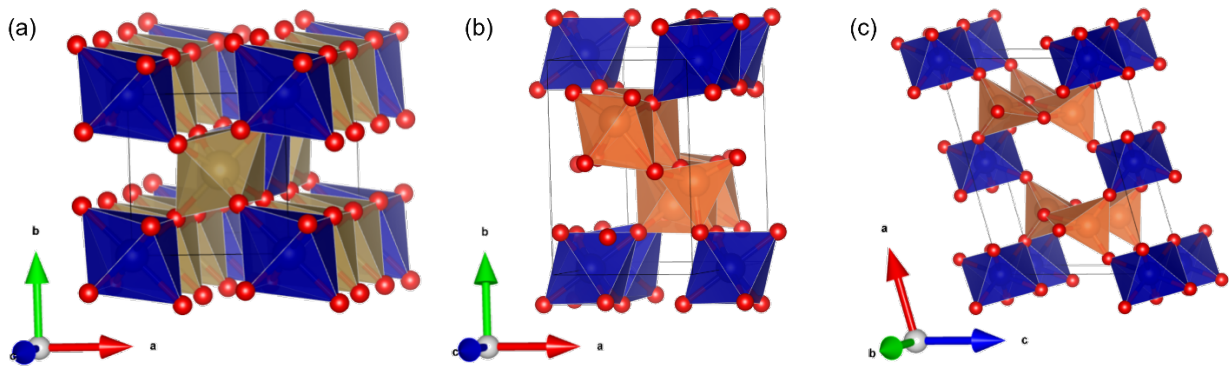

**Figure 2. Crystal structure of CoTa$_2$O$_6$ and CoV$_2$O$_6$.** The unit cell of (a) tetragonal CoTa$_2$O$_6$, (b) triclinic γ-CoV$_2$O$_6$ and (c) monoclinic α-CoV$_2$O$_6$. (Co: blue, Ta: yellow, V: orange, O: red)

*2.2 Co-containing Pyroxenes*

Pyroxenes, with a general formula AMX$_2$O$_6$ (A = mono- or di-valent cation, M = transition metal cation, X = tetravalent cation), are one of the most abundant minerals in Earth's crust. In its crystal lattice (**Fig. 3**), the edge-sharing MO$_6$ octahedra form infinite 1D chains that propagate in a zigzag fashion along the *c* axis. Adjacent chains are bridged by XO$_4$ tetrahedra along the *a* axis, while they are non-linked along the *b* axis because of the A cations residing in the cavities. This quasi-1D system can result in substantial anisotropic physical properties, hence serving as a potential platform to study in detail the emergent quantum phenomena that originate from the interplay of intra- and inter-chain couplings. The long-range antiferromagnetic unit cell of Co-containing pyroxenes was initially obtained from neutron diffraction experiments performed on the polycrystalline powder samples [44,45]. However, they were then put on the back burner because nothing particularly interesting was observed under low magnetic fields. In contrast, the analogous Cr, Mn and Fe materials have been in the spotlight due to the observed multiferroicity and magnetoelectric effects etc. [46–51]. In more recent years, the Co-containing pyroxenes have regained attention from the community due to the emergent anisotropic field-induced (meta)magnetic transitions that arise from the joint effect brought by $Co^{2+}$ ions in this highly versatile crystal structure [6,7,52,53].

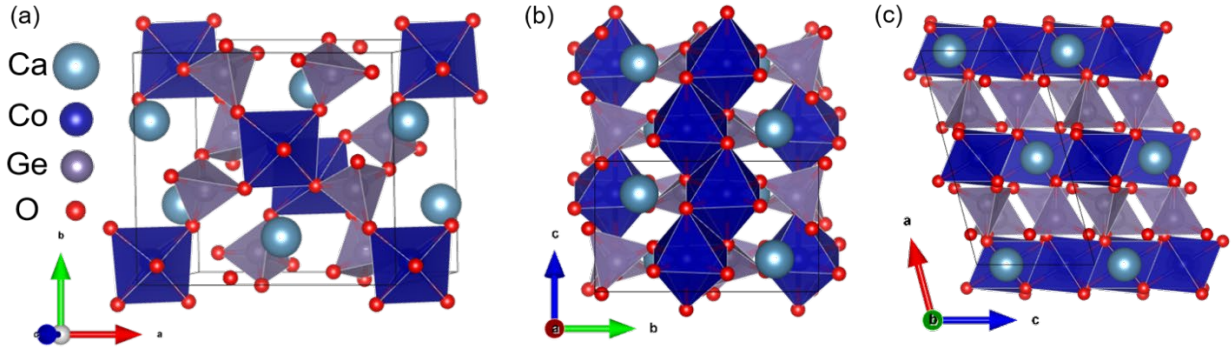

**Figure 3. Crystal structure of pyroxene ($CaCoGe_2O_6$ shown as an example).** (a) the unit cell; (b) the zig-zag 1D chain made of edge-sharing $CoO_6$ octahedra; (c) adjacent 1D chains bridged by the $GeO_4$ tetrahedra. [7]

$CoGeO_3$, a simpler version of the $ACoX_2O_6$, pyroxene, has been studied in depth. Large $CoGeO_3$ single crystals were grown in a high pressure mirror furnace. The direction-dependent magnetic characterization performed on the single crystal reveals highly anisotropic magnetic susceptibility in the system [52]. In addition, if the external magnetic field was applied along the chain direction (i.e., the *c* axis), unusually well-defined 1/3 plateaus emerged in the field-dependent magnetization data, despite the absence of an apparent triangular lattice in the structural motif [6]. (**Fig. 4**) These studies suggest that the magnetic interactions between spins in these weakly coupled Co 1D chains are more complicated than originally anticipated, hence worth further investigation.

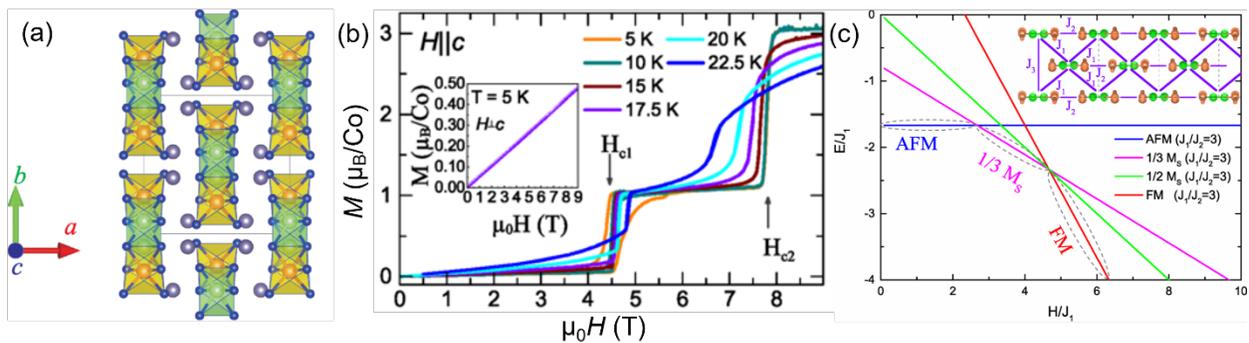

**Figure 4. CoGeO$_3$** (a) the unit cell (Green: Co1; dark yellow: Co2; blue: Oxygen; gray: Ge); (b) magnetization $M(H)$ for $H \parallel c$, inset: for $H \perp c$; (c) calculated energies for an anisotropic frustrated square lattice with one diagonal exchange interaction. [6]

The magnetic properties of the SrCoGe$_2$O$_6$ polycrystalline powder samples were studied in detail using a combination of neutron scattering, heat capacity, *ab initio* methods, and linear spin-wave theory. [53] This work demonstrates a bond-dependent Kitaev exchange model for this material, together with a field-induced state achieved through sabotaging the fragile antiferromagnetic ordering between spins. What's more, it also proposes that more intriguing phenomena might emerge once the dimensionality of the system (i.e., relative coupling strength of intra- and inter-chain interactions) has been further manipulated. [53] Following this lead, the most straightforward way to modify the dimensionality of the system is to vary the XO$_4$ bridging unit between Co 1D chains, i.e., substituting Ge with Si in the inter-chain super-exchange pathway. A systematic study of the CaCoGe$_{2-x}$Si$_x$O$_6$ series was carried out, with both end members CaCoGe$_2$O$_6$ and CaCoSi$_2$O$_6$ studied in depth by growing macroscopic single crystals. [7] On cooling below the Néel temperatures, a sharp field-induced transition in magnetization is observed for CaCoGe$_2$O$_6$, while multiple magnetization plateaus beneath the full saturation moment are spotted for CaCoSi$_2$O$_6$. These contrasting behaviors potentially arise from the different electron configurations of Ge and Si, in which the 3d orbitals are filled in the former but empty in the latter, enabling electron hopping. Thus, SiO$_4$ can aid the inter-chain super-exchange pathway between Co$^{2+}$ centers while GeO$_4$ tends to block it during magnetization. These (meta)magnetic transitions were found to be highly anisotropic, and the external magnetic field required to induce them strongly depends on the choice of bridging unit XO$_4$. (**Fig. 5**) In addition, heat capacity data collected on all these Co-containing pyroxenes revel an effective spin-½ for Co(II) ions in 1D chains, even after being manipulated by high magnetic fields. [7,53]

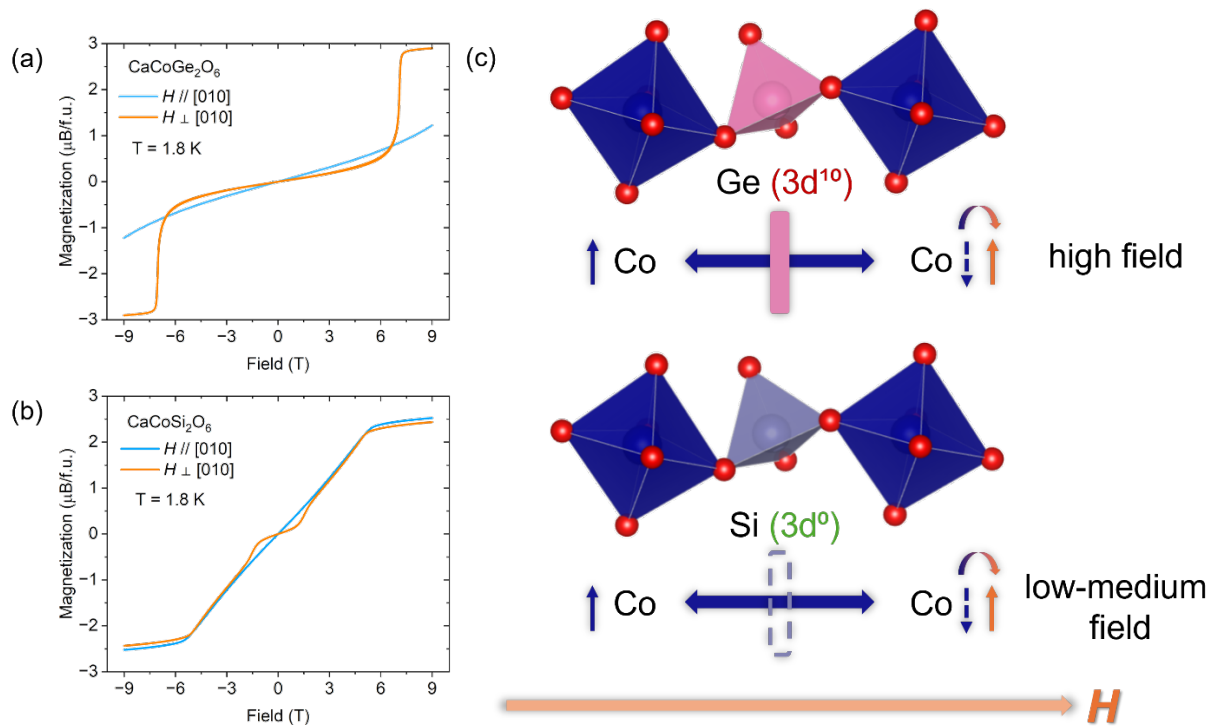

**Figure 5. CaCoXO6 pyroxene.** Direction-dependent magnetization $M$ ($H$) collected from (a) CaCoGeO$_6$ and (b) CaCoSiO$_6$; (c) A schematic illustration to show the impact of filled/empty 3d orbitals on the external field required to induce the (meta)magnetic transition. [7]

These recent studies have demonstrated that Co-containing pyroxenes can be considered as a new playground for Kitaev physics. The highly tunable dimensionality and the fruitful choice of elements within this family of materials opens more possibilities to study exotic phases with fractionalized excitations, as well as advance our understanding in Kitaev materials in general. [6,7,52,53]

*2.3 A(II)Co$_2$V$_2$O$_8$ and its derivatives*

Though the 1D chain in A(II)Co$_2$V$_2$O$_8$ family still consists of edge-sharing CoO$_6$ octahedra, it distinguishes itself from the two families discussed above by adopting a slightly different magnetic structure. Among this family, the A = Ba/Sr versions (**Fig. 6a**) have been extensively studied during the past two decades. Spins within the screw chains of Co$^{2+}$ rotating around the four-fold $c$ axis are strongly antiferromagnetically coupled, while the inter-chain coupling is much weaker (**Fig. 6b**). [54,55] This eventually yields a long-range antiferromagnetic order at substantially low temperatures (~ a few Kelvins). Thus, the A(II)Co$_2$V$_2$O$_8$ family possess a strong magnetic anisotropy, owing to the quasi-1D nature of their magnetic structure, and become effective 1D spin-1/2 antiferromagnets with Heisenberg-Ising (or XXZ) exchange anisotropy due to the $Co^{2+}$ ions in the lattice. [56] Although the screw chains in $(Ba/Sr)Co_2V_2O_8$ are antiferromagnetic, unlike the zig-zag ferromagnetic chains found in $CoNb_2O_6$, they are still rich in exotic physical phenomena including but not limited to quantum phase transitions.

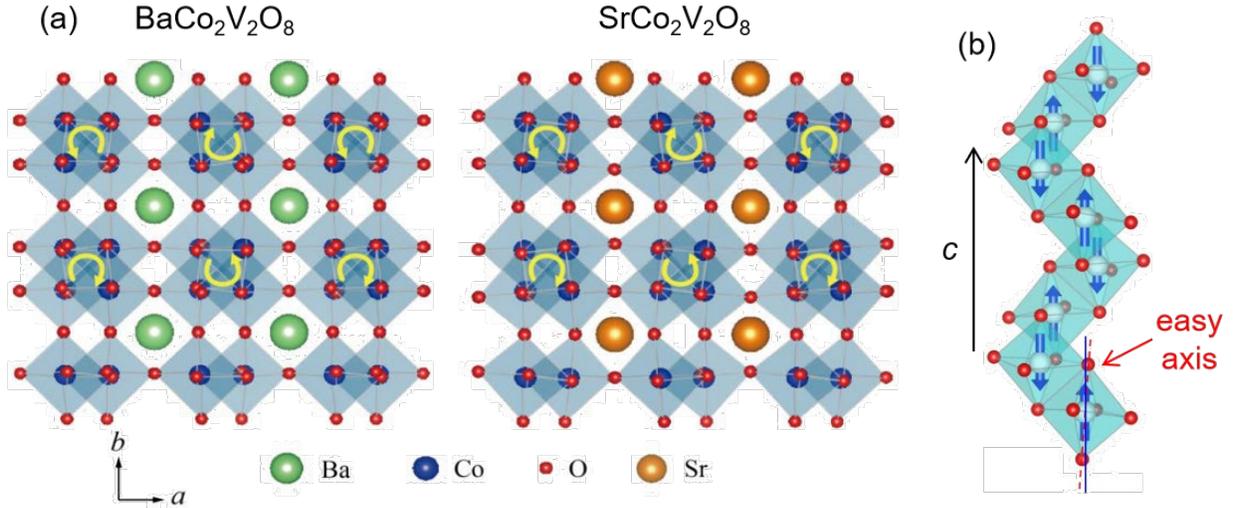

**Figure 6. The A(II)Co₂V₂O₈ family.** (a) the unit cells of $BaCo_2V_2O_8$ ($I4_1/acd$, non-polar) and $SrCo_2V_2O_8$ ($I4_1cd$, polar) viewed from the $c$ axis with nonmagnetic $VO_4$ omitted for clarity. [54] (b) the antiferromagnetically coupled spins within screw chains of $Co^{2+}$ rotating around the four-fold $c$ axis with zero applied field. [55]

Inelastic neutron scattering and terahertz spectroscopy reveal the existence of spinon confinement in $(Ba/Sr)Co_2V_2O_8$. A broad spinon continuum is observed above the Néel temperature (~ 5 K) with zero magnetic field applied. When cooling below $T_N$, pairs of spinons are confined owing to the sizable inter-chain attractive linear potential, hence the broad continuum splits into unconventional discrete spin excitations on the so-called Zeeman ladders. [57–59] The behavior of $(Ba/Sr)Co_2V_2O_8$ in an applied transverse field (perpendicular to the $c$ axis) is found to be quite intriguing as well. A substantial suppression of the antiferromagnetic order is induced by applying a transverse field along the $a$ axis in $BaCo_2V_2O_8$. However, the antiferromagnetic order can be retained even in high transverse fields, if the field is applied along the $ab$-plane diagonal. This peculiar angular-dependent phenomenon can be rationalized by the strongly anisotropic $g$-tensor for $Co^{2+}$ ions in this material, as the result of the characteristic screw chain structure. [54,56,60,61] Compared to $CoNb_2O_6$, the 1D transverse-field Ising universality has not been illustrated in $(Ba/Sr)Co_2V_2O_8$ until recently. Two QPTs in $SrCo_2V_2O_8$ are revealed in the ultra-low temperature NMR measurements, with the first QPT attributed to the rapid suppression of the antiferromagnetic order and the second one featured with gapless excitations. [62] By combining neutron diffraction and inelastic neutron scattering experiments, terahertz spectroscopy and NMR, as well as theoretical analyses, the profile of quantum phase transitions is mapped out in detail for $BaCo_2V_2O_8$, accompanied by the observation of the diagnostic $E_8$ particles. [55,63–67] The longitudinal magnetic field applied along the easy axis (approximately along the $c$ axis) further enriches the physics of this exotic family. Upon the elevated strength of an applied field, the commensurate antiferromagnetic ground state is destabilized into an incommensurate longitudinal spin density wave phase through a first-order transition, and then turns into a transverse canted antiferromagnet for both $BaCo_2V_2O_8$ [10,68] and $SrCo_2V_2O_8$ [69]. What's more spectacular is the experimental realization of Bethe strings – complex bound states of magnetic excitations in a condensed-matter system: after it was theoretically predicted by H. Bethe almost a century ago. [70] High-resolution terahertz spectroscopy [71,72] and inelastic neutron scattering [73] resolve these string states in $SrCo_2V_2O_8$ under a strong longitudinal field.

The divalent A cation does not necessarily have to be picked up from the alkaline group, for example, $Pb^{2+}$ can also reside in those cavities. Although $PbCo_2V_2O_8$ has been preliminarily investigated in the past [74,75], its detailed magnetic structure and phase diagram has not been probed until recently. A series of complex (quantum) phase transitions are discovered, regarding both the magnitude and direction of the applied magnetic field. [76] The Pb version seems to exhibit the potential for properties that could be generally comparable to $(Ba/Sr)Co_2V_2O_8$ and thus is worth further investigation. Compositions containing other divalent A cations with tolerated ionic radii, and other pentavalent cations like $Ta^{5+}$ or $Nb^{5+}$ in place of $V^{5+}$ could be the subject of future study as well.

## 3. One-dimensional Chains Consisting of Face-sharing $CoO_6$ Polyhedra
*3.1 The $A_{n+2}Co_{n+1}O_{3n+3}$ (n = 1) family*

Derived from the 2*H*-perovskite related oxides, $Ca_3Co_2O_6$ and its derivatives constitute a non-neglectable sub-class of cobalt-based quasi-1-dimensional quantum magnets and have been most intensively studied in the $A_{n+2}Co_{n+1}O_{3n+3}$ family. The crystal structure of $Ca_3Co_2O_6$ (space group *R*-3*c*) has been solved by powder X-ray and neutron diffraction data. [77] $Ca_3Co_2O_6$ distinguishes itself from other types of Co-based chain materials by the alternating arrangement of face-sharing

CoO$_6$ trigonal prisms (TP) and CoO$_6$ octahedra (Oh) within the 1D chains running along the $c$ axis (**Fig. 7a**). These chains form a triangular lattice in the $ab$ plane (**Fig. 7b**). [78]

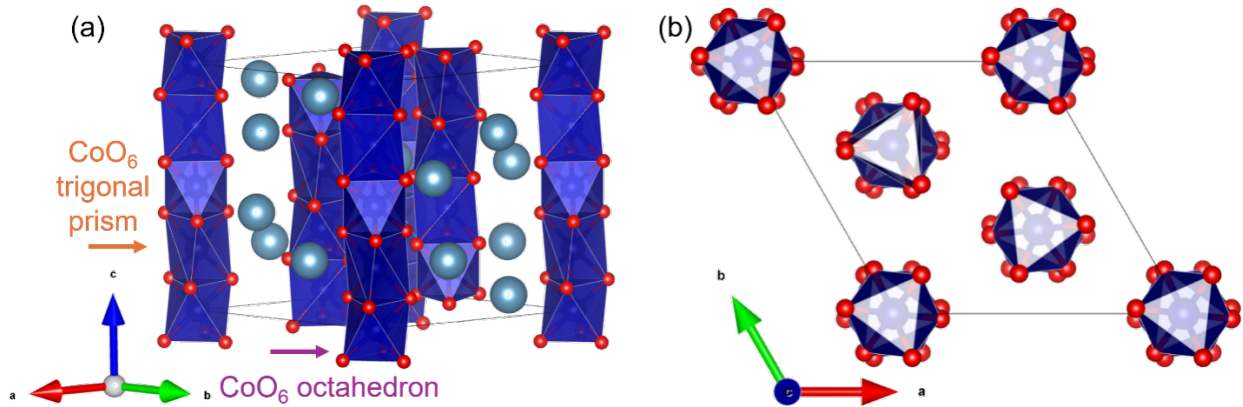

**Figure 7. Crystal structure of Ca$_3$Co$_2$O$_6$.** (a) the unit cell showing the chains running along the $c$ axis; (b) the triangular lattice formed by the chains in the $ab$ plane.

There are two types of Co centers – one resides in the trigonal prism (Co$_{trig}$) and the other resides in the octahedron (Co$_{oct}$). In order to understand the magnetism of Ca$_3$Co$_2$O$_6$, the oxidation state and electronic configuration of each type of Co center need to be mapped out first. Early experimental and theoretical studies made several suggestions, among which the two most agreed combinations are – low-spin ($S = 0$) trivalent Co$_{oct}$ and high-spin ($S = 2$) trivalent Co$_{trig}$, or low-spin (S = 1/2) tetravalent Co$_{oct}$ and high-spin (S = 3/2) divalent Co$_{trig}$. [79–86] After studying this compound in detail by using X-ray photoemission spectroscopy, X-ray absorption and magnetic circular dichroism, powder and single-crystal neutron diffraction etc., trivalent Co centers in both coordination environments (low-spin Co$_{oct}$ and high-spin Co$_{trig}$) have generally become the mainstream understanding within the community. [8,87,88]

The magnetic ground state of Ca$_3$Co$_2$O$_6$ is preliminarily depicted as a triangular lattice of ferromagnetic 1D chains (spins coupled across the spin-bearing Co$_{trig}$ sites within the chains), with the much weaker inter-chain coupling described by the antiferromagnetic Ising model. An incommensurate magnetic ground state is suggested by a few studies. [89–91] In addition to a modulated partially disordered antiferromagnetic structure along the $c$ axis, the coexistence of a shorter-range order in the $ab$ plane has also been unveiled. [8] This argument has been backed up by a magnetocaloric study of the metamagnetic transitions present in the system. [92] Field-induced plateaus are spotted in the magnetization curves of Ca$_3$Co$_2$O$_6$, which is generally rare but not rare among the subjects of this review. But what makes $Ca_3Co_2O_6$ particularly interesting is when cooling below 10 K, the plateau at 1/3 of the saturation moment (observed at the temperature range of ~ 10 - 25 K) further splits into equidistant sub-steps, signaling quantum tunneling of the magnetization. The sub-steps are clearer when a faster sweep rate is employed. [8,81,82,93–96] The origin of this intriguing phenomenon (two steps above 10 K and at least four steps below 10 K), i.e., the magnetic-order dynamics in $Ca_3Co_2O_6$ has attracted enduring interest from the community during the past two decades. Monte Carlo and Wang-Landau simulations have been carried out by theorists. Their calculation results can generally reproduce the experimental observations in quantitative agreement at different sweep rates of the applied magnetic fields. They also point out that nonequilibrium magnetization and magnetization inhomogeneity are the key factors needed to induce the four-step state below 10 K. [97–103] Formation [104] and deformation [105] of spin-density-wave microphases in $Ca_3Co_2O_6$ are investigated as well. Other than magnetic field, external factors such as time [106,107], pressure [108] and A-site chemical doping [83] have been introduced to study their influence on the magnetic order in $Ca_3Co_2O_6$. What's more, several other techniques such as orbital imaging [109], NMR [110], muon-spin spectroscopy [111] have also been used to study $Ca_3Co_2O_6$. In summary, the nature of the exotic magnetic behavior of $Ca_3Co_2O_6$ remains an ongoing puzzle and worth further investigations.

The fact that the crystal structure of $Ca_3Co_2O_6$ has two distinguished B-cation sites provides an opportunity to mix Co with other transition metals. Thus, its derivatives have also been studied by the community. The Mn-doping of $Ca_3Co_2O_6$ has drawn some attention. The discovery of collinear-magnetism-driven ferroelectricity in $Ca_3CoMnO_6$ (**Fig. 8a**) shines a spotlight on this isostructural sister compound. [112] By employing *ab initio* electronic structure calculations and X-ray absorption spectroscopy, $Co^{2+}$ and $Mn^{4+}$ adopt an ordered arrangement, with the former in trigonal prisms and the latter in octahedra, both in high-spin states. [113] The magnetic-order dynamics in the $Ca_3Co_{2-x}Mn_xO_6$ series is also examined. Similar magnetization behavior is found at much higher magnetic field sweep rates than that in $Ca_3Co_2O_6$ – six orders of magnitude faster. In addition to magnetization, steps are also spotted in the magnetostriction, electric polarization, and magnetocaloric effect, potentially proving the rich physics in this material. [114] A partially disordered antiferromagnetic phase, in $Ca_3CoRhO_6$ (**Fig. 8b**), has also been evidenced by a neutron diffraction study. [115] The oxidation states of Co and Rh have caused controversy. A spin-polarized electronic band structure calculations study suggests both cations in their trivalent states

[116], while an X-ray photoemission spectroscopy study leaning towards a combination of $Co^{2+}$ and $Rh^{4+}$ [87]. The electronic structure and/or transport properties of $Ca_3CoNiO_6$ and $Ca_3CoIrO_6$ have also been briefly studied. [87,117]

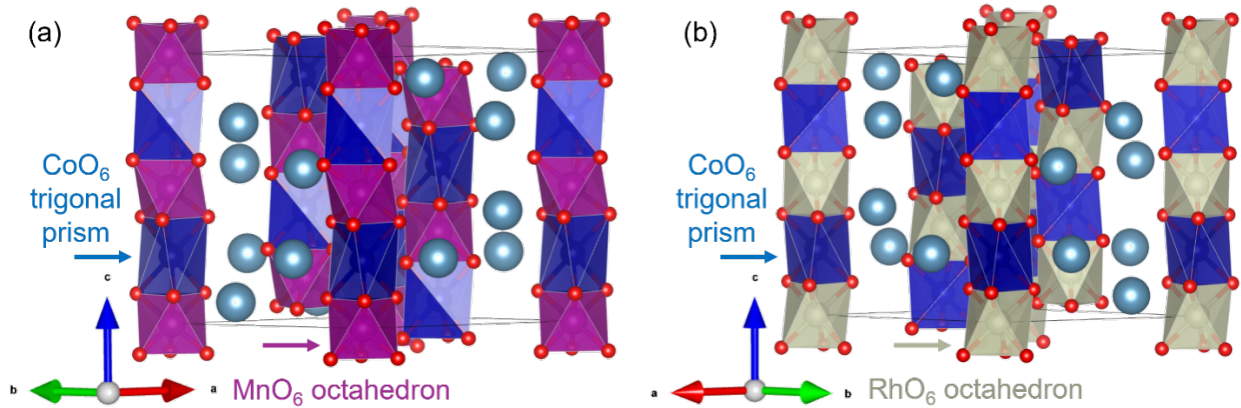

**Figure 8. Crystal structure of $Ca_3Co_2O_6$ derivatives.** The unit cell of (a) $Ca_3CoMnO_6$ and (b) $Ca_3CoRhO_6$.

*3.2 The $A_{n+2}Co_{n+1}O_{3n+3}$ (n = 2) family*

By inserting an additional octahedron into the repeating unit (1TP-1Oh) of $A_{n+2}Co_{n+1}O_{3n+3}$ (n = 1), the 1D chains running along the $c$ axis alternate in a 1TP-2Oh manner and yield a general formula of $A_{n+2}Co_{n+1}O_{3n+3}$ (n = 2). The simplest version, namely $A_4Co_3O_9$ (A = Ca, Sr, Ba), is known for its thermoelectric properties. [118] Evidence for two-dimensional antiferromagnetic order in this system has been obtained by muon spin rotation and relaxation ($\mu^+$SR) techniques. [111,119,120] However, in terms of magnetism, the Mn-doped analogues $ACo_{3-x}Mn_xO_9$ remain the main subject of study in this sub-field. The solubility of Mn in $A_4Co_3O_9$ is determined to be in the range of $0.5 \leq x \leq 2$ [121]. To avoid any potential structural disorder between $Co^{2+}$ and $Mn^{4+}$ sites, the community is particularly interested in compositions with $Co^{2+} : Mn^{4+} = 1 : 2$. [122,123] Thus, $Co^{2+}$ and $Mn^{4+}$ can adopt an ordered arrangement, with the former in trigonal prisms and the latter in octahedra (**Fig. 9a-b**). $Sr_{4-y}Ca_yCoMn_2O_9$ oxides (y = 0 and 2) are found to be an ideal playground for studying the interplay among single-ion magnetism (SIM), single-chain magnetism (SCM), and long-range order (LRO) phenomena, which are usually discussed in coordination polymers or complexes. [124] The y = 2 member has received additional attention due to its deviation from a textbook geometrically frustrated triangular lattice from the structural point of view - there are three subclasses of chains shifted from each other so some of the intrachain distances between spins differ from others. This is reflected in its *dc* magnetization data (**Fig. 9c**), two transitions denoted as $T_N$ and $T^*$ are observed in a very narrow temperature range (25 - 35 K). This double-peak feature is then further investigated by *ac* magnetization over 14 frequencies between $10^{-1}$ and $10^4$ Hz (**Fig. 9d**), revealing that a peculiar pre-transitional short-range ordering is present in the system. [5] Some other dopants like M = Zn, Cu and Rh are also explored for the $A_{n+2}Co_{n+1}O_{3n+3}$ ($n = 2$) family. [125,126]

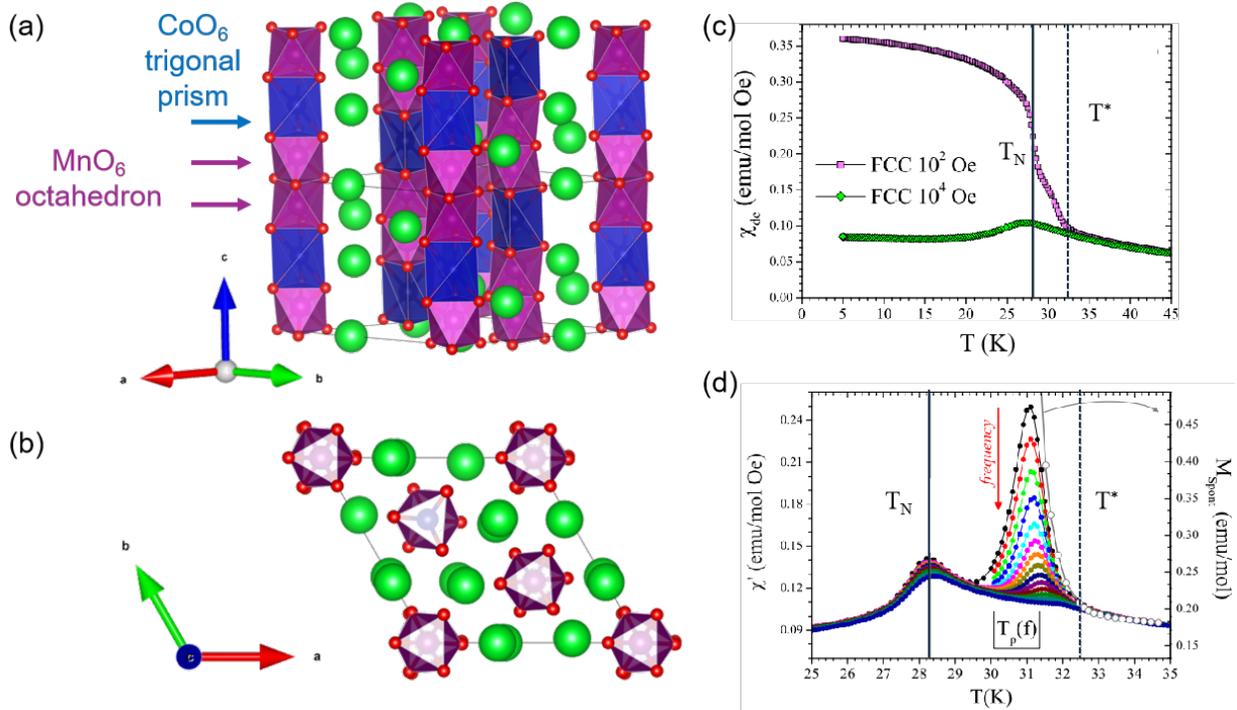

**Figure 9. The $A_4CoMn_2O_9$ family.** The unit cell of $Sr_{4-y}Ca_yCoMn_2O_9$ viewed (a) perpendicular to and (b) along the *c* axis; the (c) *dc* and (d) *ac* magnetic data collected on the $y = 2$ composition[5].

## 4. Other Co-based Quasi-1-Dimensional Quantum Magnets

*4.1 $Co^{2+}$-containing garnets*

Garnets with general composition $R_3B_2(AO_4)_3$ (R, B and A represent different individual cationic sites) have cubic symmetry and consist of a 3D network of $BO_6$ octahedra that share corners with bridging $AO_4$ tetrahedra. However, the garnet structure cannot be easily visualized in terms of close packed ionic spheres like those in perovskites or spinels, instead, it can be described as rods out of the octahedral sites that are then joined together by vacant triangular prisms. Therefore, it is only natural to anticipate that magnetic ions with strong single-ion anisotropy, such as $Co^{2+}$, can potentially lower the dimensionality of the magnetic unit cell of garnets. Consistent with this concept, by using low-temperature powder neutron diffraction, spins in $CaY_2Co_2Ge_3O_{12}$ are found to adopt an unusual anisotropic and chain-like antiferromagnetic arrangement. [127] Besides, two $Co^{2+}$ containing garnets, $CaY_2Co_2Ge_3O_{12}$ and $NaCa_2Co_2V_3O_{12}$, are reported to exhibit magnetic-field-induced dielectric anomalies. [128] Garnets can accommodate fruitful choices of elements, and thus $Co^{2+}$-containing garnets might be a class of materials that have been overlooked in terms of low-dimensional magnetism due to its apparent highly symmetric cubic crystal structure.

*4.2 $Co^{2+}$-containing halide compounds*

$CsCoX_3$ (X = Cl and Br) are reported to exhibit exotic phenomena that are analogous to some of the oxides discussed above in this review. Raman scattering measurements have been performed to study the spin dynamics in this system. [129–132] The spin-wave response corresponding to an $S = ½$ Ising-like chain antiferromagnet is unveiled by inelastic neutron scattering experiments [133], as well as the domain walls [134,135] previously predicted by Villain [136] and soliton response occurring at elevated temperatures [137]. Theorists have also made multiple attempts to justify these experimental observations [138–141].

Although $RbCoCl_3$ has been preliminarily investigated by Fourier Transform infrared spectroscopy [142] and Raman scattering [132,143], it regains some attention from the community very recently. By combining the high-resolution neutron spectroscopy with Monte Carlo simulations, the thermal control of spin excitations in $RbCoCl_3$ **(Fig. 10a-b)** is investigated to an unprecedented extent [144]. This work motivates a few more studies to further characterize the two magnetic phase transitions at low temperatures and the unusual behavior of the spin-phonon coupling in $RbCoCl_3$. [145–147]

*4.3 $Co^{2+}$-containing hybrid compounds*

Hybrid compounds including both organic and inorganic building blocks could offer additional degrees of freedom when modifying the structural motif to tailor the need for target properties. The size of organic building blocks varies in a much wider range than is typically seen for inorganics, thus, the dimensionality of the $Co^{2+}$-containing hybrid compounds could be tuned to a greater extent compared to those inorganic ones. A few examples of Co-based one-dimensional spin-chain hybrid compounds have been studied by the community, displaying the merit of the versatility of organic ligands.

For example, neutron scattering measurements along with magnetic characterizations reveal that the dominant interaction in the $Co^{2+}$-containing hydrazinium sulfate material $Co(N_2H_5)_2(SO_4)_2$ (**Fig. 10c**) is the intra-chain antiferromagnetic interaction. Polarized neutron powder diffraction measurements also point out that this material exhibits an anisotropic magnetization on the Co site with indications of Ising-type interactions in the *ab* plane, thus motivating the need for the growth of large, high-quality single crystals to further study the direction-dependent physical properties in this system. [148] In addition, the reaction of $Co(NCS)_2$ with 3-bromopyridine leads to the formation of various types of complexes depending on the solvent and reaction conditions used. This series of complexes show qualitatively different magnetic properties upon variation in the coordination environment of $Co^{2+}$ ions and the overall dimensionality of materials. [149] These reported $Co^{2+}$-containing hybrid compounds demonstrate that this is a sub-class of quasi-1D chain materials with high potential in quantum magnetism.

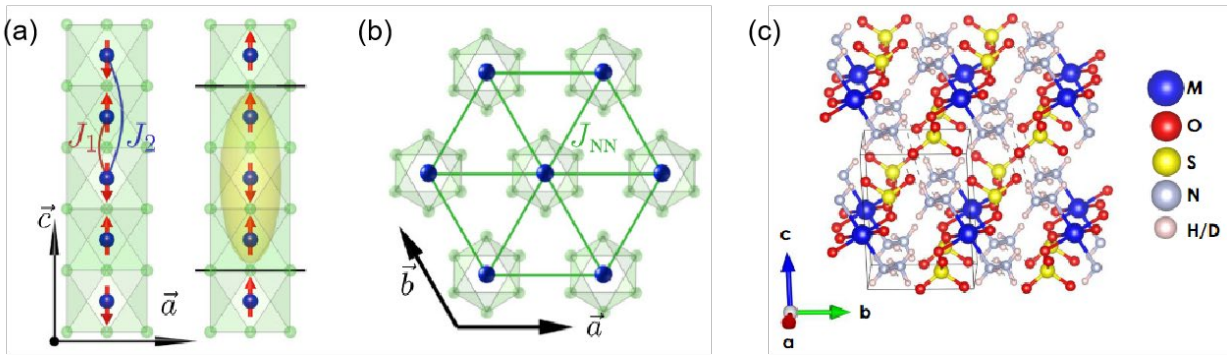

**Figure 10. Crystal structure of $Co^{2+}$-containing halide and hybrid compounds.** (a) schematic representation of chains in $RbCoCl_3$, with $J_1$ and $J_2$ denoting the magnetic coupling strength between nearest-neighboring and next-nearest-neighboring $Co^{2+}$ centers, and the yellow ellipse denoting a pair of interacting domain walls; [144] (b) the triangular lattice of chains in $RbCoCl_3$ with $J_{NN}$ denoting the magnetic coupling strength between nearest-neighboring $Co^{2+}$ centers in the *ab* plane; [144] (c) crystal structure of $Co(N_2H_5)_2(SO_4)_2$. [148]

*4.4 $Co^{2+}$-containing chalcogenides*

For quasi-1D $Co^{2+}$-containing chalcogenides, these systems warrant a separate review, as the different bonding interactions associated with oxide and heavier chalcogenide anions lead to distinct yet equally fascinating magnetic behaviors in the quantum regime.

## 5. Conclusions

In this topical review, we have summarized many of the intensively studied Co-based quasi-1-dimensional materials that are known or proposed to be quantum magnets over the past few decades, ranging from quasi-1D spin-chain compounds consisting of edge-sharing and face-sharing $CoO_6$ polyhedra, to other $Co^{2+}$-containing systems such as highly symmetric garnets, halide and hybrid compounds. The exotic phenomena observed either commonly or unique among the subjects of the current review, make them one of the keen research topics in both the theoretical and experimental physics world. While some of the problems have been figured out, due to the development of modern theories, characterization techniques and advances in sample preparation skills, there are still a lot of open questions under ongoing debate, affirming Co-based quasi-1-dimensional materials an active playground with huge potential for studying magnetism in the quantum regime.

Among the subjects of current review, we can find that some material systems such as pyroxenes, garnets etc. are not even studied in an equivalent depth compared to the others, not to mention that some of the studies only came in very recently. Besides, the material systems intensively studied in other keen research areas of condensed matter physics substantially outnumber those in this one. Hence it is of priority to overcome obstacles in high-quality crystal growth and enrich the variety of material systems investigated in this field. Therefore, by writing this review, we hope to bring researchers with comprehensive understanding of both condensed-matter physics and materials chemistry on board. Because this is an exciting interdisciplinary research field that should attract scientists who work at the border between physics and chemistry.

## Acknowledgements

This work was supported by the Start-up Research Fund of Southeast University, Grant No. RF1028624196, by the Gordon and Betty Moore foundation, EPiQS initiative, Grant GBMF-9066, and the Basic Sciences Division of the US Department of Energy, grant DE-FG02-98ER45706. L.J. acknowledges the support of the open research fund of Key Laboratory of Quantum Materials and Devices of Ministry of Education (Southeast University).

## Competing Interests

The authors declare no competing interests.

## References

[1] S. Lee, R.K. Kaul, L. Balents, Interplay of quantum criticality and geometric frustration in columbite, Nat Phys 6 (2010) 702–706. https://doi.org/10.1038/nphys1696.

[2] T. Liang, S.M. Koohpayeh, J.W. Krizan, T.M. McQueen, R.J. Cava, N.P. Ong, Heat capacity peak at the quantum critical point of the transverse Ising magnet $CoNb_2O_6$, Nat Commun 6 (2015) 7611. https://doi.org/10.1038/ncomms8611.

[3] G.H. Wannier, The Triangular Ising Net, Physical Review 79 (1950) 357–364. https://doi.org/10.1103/PhysRev.79.357.

[4] W. Scharf, H. Weitzel, I. Yaeger, I. Maartense, B.M. Wanklyn, Magnetic Structures of $CoNb_2O_6$, J Magn Magn Mater 13 (1979) 121–124. https://doi.org/10.1016/0304-8853(79)90044-1.

[5] V. Hardy, V. Caignaert, O. Pérez, L. Hervé, N. Sakly, B. Raveau, M.M. Seikh, F. Damay, Pretransitional short-range ordering in a triangular lattice of Ising spin chains, Phys Rev B 98 (2018) 144414. https://doi.org/10.1103/PhysRevB.98.144414.

[6] H. Guo, L. Zhao, M. Baenitz, X. Fabrèges, A. Gukasov, A. Melendez Sans, D.I. Khomskii, L.H. Tjeng, A.C. Komarek, Emergent 1/3 magnetization plateaus in pyroxene $CoGeO_3$, Phys Rev Res 3 (2021) L032037. https://doi.org/10.1103/PhysRevResearch.3.L032037.

[7] L. Jin, S. Peng, A. Nakano Rutherford, X. Xu, D. Ni, C. Yang, Y. Ji Byeon, W. Xie, H. Zhou, X. Dai, R.J. Cava, A pyroxene-based quantum magnet with multiple magnetization plateaus, Sci Adv 10 (2024) eadp4685. https://doi.org/10.1126/sciadv.adp4685.

[8] S. Agrestini, L.C. Chapon, A. Daoud-Aladine, J. Schefer, A. Gukasov, C. Mazzoli, M.R. Lees, O.A. Petrenko, Nature of the magnetic order in $Ca_3Co_2O_6$, Phys Rev Lett 101 (2008) 097207. https://doi.org/10.1103/PhysRevLett.101.097207.

[9] S. Kobayashi, S. Mitsuda, K. Prokes, Low-temperature magnetic phase transitions of the geometrically frustrated isosceles triangular Ising antiferromagnet $CoNb_2O_6$, Phys Rev B 63 (2000) 024415. https://doi.org/10.1103/PhysRevB.63.024415.

[10] E. Canévet, B. Grenier, M. Klanjšek, C. Berthier, M. Horvatić, V. Simonet, P. Lejay, Field-induced magnetic behavior in quasi-one-dimensional Ising-like antiferromagnet $BaCo_2V_2O_8$: A single-crystal neutron diffraction study, Phys Rev B 87 (2013) 054408. https://doi.org/10.1103/PhysRevB.87.054408.

[11] A.B. Zamolodchikov, Integrals of motion and S-matrix of the (scaled) T=Tc Ising model with magnetic field, Int J Mod Phys A 4 (1989) 4235–4248. https://doi.org/10.1142/S0217751X8900176X.

[12] S. Sachdev, Physics World Quantum phase transitions, Cambridge Univ. Press, 1999.

[13] S. Ghosh, T.F. Rosenbaum, G. Aeppli, S.N. Coppersmith, Entangled quantum state of magnetic dipoles, Nature 425 (2003) 48–51. https://doi.org/10.1038/nature01888.

[14] R. Coldea, D.A. Tennant, E.M. Wheeler, E. Wawrzynska, D. Prabhakaran, M. Telling, K. Habicht, P. Smeibidl, K. Kiefer, Quantum Criticality in an Ising Chain: Experimental Evidence for Emergent $E_8$ Symmetry, Science 327 (2010) 177–180. https://doi.org/10.1126/science.1180085.

[15] E. Lier, T. Schultz, D.J. MATTE Thomas Watson, Two Soluble Models of an Antiferromagnetic Chain, Ann Phys (NY) 16 (1961) 407–466.

[16] T. Hanawa, K. Shinkawa, M. Ishikawa, K. Miyatani, K. Saito, K. Kohn, Anisotropic specific heat of $CoNb_2O_6$ in magnetic fields, J Physical Soc Japan 63 (1994) 2706–2715. https://doi.org/https://doi.org/10.1143/JPSJ.63.2706.

[17] C. Heid, H. Weitzel, B.M. Bonnet W Gonschorek, T.J. Vogt Norwig, H. Fuess, Magnetic phase diagram of $CoNb_2O_6$: a neutron diffraction study, J Magn Magn Mater 151 (1995) 123–131. https://doi.org/10.1016/0304-8853(95)00394-0.

[18] S. Kobayashi, S. Mitsuda A', K. Hosoya, H. Yoshizawa, T. Hanawa, M. Ishikawa, K. Miyatani, K. Saito, K. Kohn, Competition between the inter-chain interaction and single-ion anisotropy in $CoNb_2O_6$, Physica B Condens Matter 213 (1995) 176–178. https://doi.org/10.1016/0921-4526(95)00096-R.

[19] S. Kobayashi, S. Mitsuda, M. Ishikawa, K. Miyatani, K. Kohn, Three-dimensional magnetic ordering in the quasi-one-dimensional Ising magnet $CoNb_2O_6$ with partially released geometrical frustration, Phys Rev B 60 (1999) 3331. https://doi.org/10.1103/PhysRevB.60.3331.

[20] A.W. Kinross, M. Fu, T.J. Munsie, H.A. Dabkowska, G.M. Luke, S. Sachdev, T. Imai, Evolution of quantum fluctuations near the quantum critical point of the transverse field ising chain system $CoNb_2O_6$, Phys Rev X 4 (2014) 031008. https://doi.org/10.1103/PhysRevX.4.031008.

[21] K. Matsuura, P.T. Cong, S. Zherlitsyn, J. Wosnitza, N. Abe, T.H. Arima, Anomalous Lattice Softening Near a Quantum Critical Point in a Transverse Ising Magnet, Phys Rev Lett 124 (2020) 127205. https://doi.org/10.1103/PhysRevLett.124.127205.

[22] C.M. Morris, R. Valdés Aguilar, A. Ghosh, S.M. Koohpayeh, J. Krizan, R.J. Cava, O. Tchernyshyov, T.M. McQueen, N.P. Armitage, Hierarchy of bound states in the one-dimensional ferromagnetic Ising Chain $CoNb_2O_6$ investigated by high-resolution time-domain terahertz spectroscopy, Phys Rev Lett 112 (2014) 137403. https://doi.org/10.1103/PhysRevLett.112.137403.

[23] J.A. Kjäll, F. Pollmann, J.E. Moore, Bound states and E8 symmetry effects in perturbed quantum Ising chains, Phys Rev B 83 (2011) 020407(R). https://doi.org/10.1103/PhysRevB.83.020407.

[24] N.J. Robinson, F.H.L. Essler, I. Cabrera, R. Coldea, Quasiparticle breakdown in the quasi-one-dimensional Ising ferromagnet $CoNb_2O_6$, Phys Rev B 90 (2014) 174406. https://doi.org/10.1103/PhysRevB.90.174406.

[25] M. Fava, R. Coldea, S.A. Parameswaran, Glide symmetry breaking and Ising criticality in the quasi-1D magnet $CoNb_2O_6$, Proc Natl Acad Sci U S A 117 (2020) 25219–25224. https://doi.org/10.1073/pnas.2007986117.

[26] C.L. Sarkis, S. Säubert, V. Williams, E.S. Choi, T.R. Reeder, H.S. Nair, K.A. Ross, Low-temperature domain-wall freezing and nonequilibrium dynamics in the transverse-field Ising model material $CoNb_2O_6$, Phys Rev B 104 (2021) 214424. https://doi.org/10.1103/PhysRevB.104.214424.

[27] L. Woodland, D. Macdougal, I.M. Cabrera, J.D. Thompson, D. Prabhakaran, R.I. Bewley, R. Coldea, Tuning the confinement potential between spinons in the Ising chain compound $CoNb_2O_6$ using longitudinal fields and quantitative determination of the microscopic Hamiltonian, Phys Rev B 108 (2023) 184416. https://doi.org/10.1103/PhysRevB.108.184416.

[28] S. Birnkammer, J. Knolle, M. Knap, Signatures of Domain-Wall Confinement in Raman Spectroscopy of Ising Spin Chains, Phys Rev B 110 (2024) 134408. https://doi.org/10.1103/PhysRevB.110.134408.

[29] C.M. Morris, N. Desai, J. Viirok, D. Hüvonen, U. Nagel, T. Rõõm, J.W. Krizan, R.J. Cava, T.M. McQueen, S.M. Koohpayeh, R.K. Kaul, N.P. Armitage, Duality and domain wall dynamics in a twisted Kitaev chain, Nat Phys 17 (2021) 832–836. https://doi.org/10.1038/s41567-021-01208-0.

[30] D. Churchill, H.Y. Kee, Transforming from Kitaev to Disguised Ising Chain: Application to $CoNb_2O_6$, Phys Rev Lett 133 (2024) 056703. https://doi.org/10.1103/PhysRevLett.133.056703.

[31] J.N. Reimers, J.E. Greedan, C. V Stager, R. Kremer, Crystal Structure and Magnetism in $CoSb_2O_6$ and $CoTa_2O_6$, J Solid State Chem 83 (1989) 20–30. https://doi.org/10.1016/0022-4596(89)90049-2.

[32] I.S. Mulla, N. Natarajan, A.B. Gaikwad, V. Samuel, U.N. Guptha, V. Ravi, A coprecipitation technique to prepare $CoTa_2O_6$ and $CoNb_2O_6$, Mater Lett 61 (2007) 2127–2129. https://doi.org/10.1016/j.matlet.2006.08.024.

[33] E.J. Kinast, C.A. dos Santos, D. Schmitt, O. Isnard, M.A. Gusmão, J.B.M. da Cunha, Magnetic structure of the quasi-two-dimensional compound $CoTa_2O_6$, J Alloys Compd 491 (2010) 41–44. https://doi.org/10.1016/j.jallcom.2009.10.241.

[34] A.B. Christian, A.T. Schye, K.O. White, J.J. Neumeier, Magnetic, thermal, and optical properties of single-crystalline $CoTa_2O_6$ and $FeTa_2O_6$ and their anisotropic magnetocaloric effect, J Condens Matter Phys 30 (2018) 195803. https://doi.org/10.1088/1361-648X/aab884.

[35] Z. He, M. Itoh, Single crystal flux growth of the Ising spin-chain system $\gamma$-$CoV_2O_6$, J Cryst Growth 388 (2014) 103–106. https://doi.org/10.1016/j.jcrysgro.2013.11.078.

[36] S.A.J. Kimber, H. Mutka, T. Chatterji, T. Hofmann, P.F. Henry, H.N. Bordallo, D.N. Argyriou, J.P. Attfield, Metamagnetism and soliton excitations in the modulated ferromagnetic Ising chain $CoV_2O_6$, Phys Rev B 84 (2011) 104425. https://doi.org/10.1103/PhysRevB.84.104425.

[37] M. Markkula, A.M. Arévalo-López, J.P. Attfield, Field-induced spin orders in monoclinic $CoV_2O_6$, Phys Rev B 86 (2012) 134401. https://doi.org/10.1103/PhysRevB.86.134401.

[38] M. Markkula, A.M. Arevalo-Lopez, J.P. Attfield, Neutron diffraction study of monoclinic brannerite-type $CoV_2O_6$, J Solid State Chem 192 (2012) 390–393. https://doi.org/10.1016/j.jssc.2012.04.029.

[39] M. Lenertz, J. Alaria, D. Stoeffler, S. Colis, A. Dinia, O. Mentré, G. André, F. Porcher, E. Suard, Magnetic structure of ground and field-induced ordered states of low-dimensional $\alpha$-$CoV_2O_6$: Experiment and theory, Phys Rev B 86 (2012) 214428. https://doi.org/10.1103/PhysRevB.86.214428.

[40] A. Saúl, D. Vodenicarevic, G. Radtke, Theoretical study of the magnetic order in $\alpha$-$CoV_2O_6$, Phys Rev B 87 (2013) 024403. https://doi.org/10.1103/PhysRevB.87.024403.

[41] Z. He, J.I. Yamaura, Y. Ueda, W. Cheng, $CoV_2O_6$ single crystals grown in a closed crucible: Unusual magnetic behaviors with large anisotropy and 1/3 magnetization plateau, J Am Chem Soc 131 (2009) 7554–7555. https://doi.org/10.1021/ja902623b.

[42] Y. Drees, S. Agrestini, O. Zaharko, A.C. Komarek, Floating zone single crystal growth of $\gamma$-$CoV_2O_6$ with substantially enhanced crystal size and quality, Cryst Growth Des 15 (2015) 1168–1172. https://doi.org/10.1021/cg5015303.

[43] Y. Zhao, Z. Ma, Z. He, H. Liao, Y.C. Wang, J. Wang, Y. Li, Quantum annealing of a frustrated magnet, Nat Commun 15 (2024) 3495. https://doi.org/10.1038/s41467-024-47819-y.

[44] G. Durand, S. Vilminot, P. Rabu, A. Derory, J.P. Lambour, E. Ressouche, Synthesis, Structure, and Magnetic Properties of CaMSi$_2$O$_6$(M= Co, Ni) Compounds and Their Solid Solutions, J Solid State Chem 124 (1996) 374–380. https://doi.org/10.1006/JSSC.1996.0252.

[45] G.J. Redhammer, G. Roth, W. Treutmann, W. Paulus, G. André, C. Pietzonka, G. Amthauer, Magnetic ordering and spin structure in Ca-bearing clinopyroxenes CaM$^{2+}$(Si, Ge)$_2$O$_6$, M=Fe, Ni, Co, Mn, J Solid State Chem 181 (2008) 3163–3176. https://doi.org/10.1016/j.jssc.2008.08.014.

[46] S. Jodlauk, P. Becker, J.A. Mydosh, D.I. Khomskii, T. Lorenz, S. V Streltsov, D.C. Hezel, L. Bohatý, Pyroxenes: a new class of multiferroics, J Condens Matter Phys 19 (2007) 432201. https://doi.org/10.1088/0953-8984/19/43/432201.

[47] I. Kim, B.-G. Jeon, D. Patil, S. Patil, G. Nénert, K.H. Kim, Observation of multiferroic properties in pyroxene NaFeGe$_2$O$_6$, J Condens Matter Phys 24 (2012) 306001. https://doi.org/10.1088/0953-8984/24/30/306001.

[48] M. Ackermann, L. Andersen, T. Lorenz, L. Bohatý, P. Becker, Anisotropy study of multiferroicity in the pyroxene NaFeGe$_2$O$_6$, New J Phys 17 (2015) 013045. https://doi.org/10.1088/1367-2630/17/1/013045.

[49] L. Ding, C. V. Colin, C. Darie, J. Robert, F. Gay, P. Bordet, One-dimensional short-range magnetic correlations in the magnetoelectric pyroxene CaMnGe$_2$O$_6$, Phys Rev B 93 (2016). https://doi.org/10.1103/PhysRevB.93.064423.

[50] G.J. Redhammer, G. Roth, A. Senyshyn, G. Tippelt, C. Pietzonka, Crystal and magnetic spin structure of Germanium-Hedenbergite, CaFeGe$_2$O$_6$, and a comparison with other magnetic/magnetoelectric/ multiferroic pyroxenes, Zeitschrift Fur Kristallographie 228 (2013) 140–150. https://doi.org/10.1524/zkri.2013.1586.

[51] J. O'Connell, X. Xu, L. Jin, R.J. Cava, The ferromagnetic to antiferromagnetic crossover in chromium-based pyroxenes tuned via the Ge to Si ratio, J Solid State Chem 339 (2024) 124919. https://doi.org/10.1016/j.jssc.2024.124919.

[52] L. Zhao, Z. Hu, H. Guo, C. Geibel, H.J. Lin, C. Te Chen, D. Khomskii, L.H. Tjeng, A.C. Komarek, Single crystal growth and physical properties of pyroxene CoGeO$_3$, Crystals (Basel) 11 (2021) 378. https://doi.org/10.3390/cryst11040378.

[53] P.A. Maksimov, A. V. Ushakov, A.F. Gubkin, G.J. Redhammer, S.M. Winter, A.I. Kolesnikov, A.M. dos Santos, Z. Gai, M.A. McGuire, A. Podlesnyak, S. V. Streltsov, Cobalt-based pyroxenes: A new playground for Kitaev physics, Proc Natl Acad Sci U S A 121 (2024) e2409154121. https://doi.org/10.1073/pnas.2409154121.

[54] A. Okutani, T. Kida, T. Usui, T. Kimura, K. Okunishi, M. Hagiwara, High Field Magnetization of Single Crystals of the S=1/2 Quasi-1D Ising-like Antiferromagnet SrCo$_2$V$_2$O$_8$, in: Phys Procedia, Elsevier B.V., 2015: pp. 779–784. https://doi.org/10.1016/j.phpro.2015.12.101.

[55] Q. Faure, S. Takayoshi, S. Petit, V. Simonet, S. Raymond, L.P. Regnault, M. Boehm, J.S. White, M. Månsson, C. Rüegg, P. Lejay, B. Canals, T. Lorenz, S.C. Furuya, T. Giamarchi, B. Grenier, Topological quantum phase transition in the Ising-like antiferromagnetic spin chain BaCo$_2$V$_2$O$_8$, Nat Phys 14 (2018) 716–722. https://doi.org/10.1038/s41567-018-0126-8.

[56] S. Kimura, H. Yashiro, M. Hagiwara, K. Okunishi, K. Kindo, Z. He, T. Taniyama, M. Itoh, High field magnetism of the quasi one-dimensional anisotropic antiferromagnet BaCo$_2$V$_2$O$_8$, in: J Phys Conf Ser, Institute of Physics Publishing, 2006: pp. 99–102. https://doi.org/10.1088/1742-6596/51/1/021.

[57] B. Grenier, S. Petit, V. Simonet, E. Canévet, L.P. Regnault, S. Raymond, B. Canals, C. Berthier, P. Lejay, Longitudinal and transverse Zeeman ladders in the Ising-like chain antiferromagnet BaCo$_2$V$_2$O$_8$, Phys Rev Lett 114 (2015) 017201. https://doi.org/10.1103/PhysRevLett.114.017201.

[58] Z. Wang, M. Schmidt, A.K. Bera, A.T.M.N. Islam, B. Lake, A. Loidl, J. Deisenhofer, Spinon confinement in the one-dimensional Ising-like antiferromagnet SrCo$_2$V$_2$O$_8$, Phys Rev B 91 (2015) 140404(R). https://doi.org/10.1103/PhysRevB.91.140404.

[59] A.K. Bera, B. Lake, F.H.L. Essler, L. Vanderstraeten, C. Hubig, U. Schollwöck, A.T.M.N. Islam, A. Schneidewind, D.L. Quintero-Castro, Spinon confinement in a quasi-one-dimensional anisotropic Heisenberg magnet, Phys Rev B 96 (2017) 054423. https://doi.org/10.1103/PhysRevB.96.054423.

[60] S. Kimura, K. Okunishi, M. Hagiwara, K. Kindo, Z. He, T. Taniyama, M. Itoh, K. Koyama, K. Watanabe, Collapse of magnetic order of the quasi one-dimensional ising-like antiferromagnet BaCo$_2$V$_2$O$_8$ in transverse fields, J Physical Soc Japan 82 (2013) 033706. https://doi.org/10.7566/JPSJ.82.033706.

[61] S.K. Niesen, G. Kolland, M. Seher, O. Breunig, M. Valldor, M. Braden, B. Grenier, T. Lorenz, Magnetic phase diagrams, domain switching, and quantum phase transition of the quasi-one-dimensional Ising-like antiferromagnet BaCo$_2$V$_2$O$_8$, Phys Rev B 87 (2013) 224413. https://doi.org/10.1103/PhysRevB.87.224413.

[62] Y. Cui, H. Zou, N. Xi, Z. He, Y.X. Yang, L. Shu, G.H. Zhang, Z. Hu, T. Chen, R. Yu, J. Wu, W. Yu, Quantum Criticality of the Ising-like Screw Chain Antiferromagnet SrCo$_2$V$_2$O$_8$ in a Transverse Magnetic Field, Phys Rev Lett 123 (2019) 067203. https://doi.org/10.1103/PhysRevLett.123.067203.

[63] Q. Faure, S. Takayoshi, B. Grenier, S. Petit, S. Raymond, M. Boehm, P. Lejay, T. Giamarchi, V. Simonet, Solitonic excitations in the Ising anisotropic chain BaCo$_2$V$_2$O$_8$ under large transverse

magnetic field, Phys Rev Res 3 (2021) 043227.
https://doi.org/10.1103/PhysRevResearch.3.043227.

[64] Z. Zhang, K. Amelin, X. Wang, H. Zou, J. Yang, U. Nagel, T. Roõ&tild;m, T. Dey, A.A. Nugroho, T. Lorenz, J. Wu, Z. Wang, Observation of $E_8$ particles in an Ising chain antiferromagnet, Phys Rev B 101 (2020) 220411(R). https://doi.org/10.1103/PhysRevB.101.220411.

[65] X. Wang, K. Puzniak, K. Schmalzl, C. Balz, M. Matsuda, A. Okutani, M. Hagiwara, J. Ma, J. Wu, B. Lake, Spin dynamics of the $E_8$ particles, Sci Bull (Beijing) 69 (2024) 2974–2977. https://doi.org/10.1016/j.scib.2024.07.040.

[66] H. Zou, Y. Cui, X. Wang, Z. Zhang, J. Yang, G. Xu, A. Okutani, M. Hagiwara, M. Matsuda, G. Wang, G. Mussardo, K. Hódsági, M. Kormos, Z. He, S. Kimura, R. Yu, W. Yu, J. Ma, J. Wu, E8 Spectra of Quasi-One-Dimensional Antiferromagnet $BaCo_2V_2O_8$ under Transverse Field, Phys Rev Lett 127 (2021) 077201. https://doi.org/10.1103/PhysRevLett.127.077201.

[67] X. Wang, H. Zou, K. Hódsági, M. Kormos, G. Takács, J. Wu, Cascade of singularities in the spin dynamics of a perturbed quantum critical Ising chain, Phys Rev B 103 (2021) 235117. https://doi.org/10.1103/PhysRevB.103.235117.

[68] B. Grenier, V. Simonet, B. Canals, P. Lejay, M. Klanjšek, M. Horvatić, C. Berthier, Neutron diffraction investigation of the H-T phase diagram above the longitudinal incommensurate phase of $BaCo_2V_2O_8$, Phys Rev B 92 (2015) 134416. https://doi.org/10.1103/PhysRevB.92.134416.

[69] L. Shen, O. Zaharko, J.O. Birk, E. Jellyman, Z. He, E. Blackburn, Magnetic phase diagram of the quantum spin chain compound $SrCo_2V_2O_8$: A single-crystal neutron diffraction study, New J Phys 21 (2019) 073014. https://doi.org/10.1088/1367-2630/ab2b7a.

[70] H. Von Bethe, Zur Theorie der Metalle, Zeitschrift Für Physik 71 (1931) 205–226.

[71] Z. Wang, J. Wu, W. Yang, A.K. Bera, D. Kamenskyi, A.T.M.N. Islam, S. Xu, J.M. Law, B. Lake, C. Wu, A. Loidl, Experimental observation of Bethe strings, Nature 554 (2018) 219–223. https://doi.org/10.1038/nature25466.

[72] Z. Wang, M. Schmidt, A. Loidl, J. Wu, H. Zou, W. Yang, C. Dong, Y. Kohama, K. Kindo, D.I. Gorbunov, S. Niesen, O. Breunig, J. Engelmayer, T. Lorenz, Quantum Critical Dynamics of a Heisenberg-Ising Chain in a Longitudinal Field: Many-Body Strings versus Fractional Excitations, Phys Rev Lett 123 (2019) 067202. https://doi.org/10.1103/PhysRevLett.123.067202.

[73] A.K. Bera, J. Wu, W. Yang, R. Bewley, M. Boehm, J. Xu, M. Bartkowiak, O. Prokhnenko, B. Klemke, A.T.M.N. Islam, J.M. Law, Z. Wang, B. Lake, Dispersions of many-body Bethe strings, Nat Phys 16 (2020) 625–630. https://doi.org/10.1038/s41567-020-0835-7.

[74] Z. He, Y. Ueda, M. Itoh, Synthesis, structure and magnetic properties of new vanadate $PbCo_2V_2O_8$, J Solid State Chem 180 (2007) 1770–1774. https://doi.org/10.1016/j.jssc.2007.03.026.

[75] Z. He, Y. Ueda, M. Itoh, Field-induced order-disorder transition in quasi-one-dimensional spin system $PbCo_2V_2O_8$, Solid State Commun 142 (2007) 404–406. https://doi.org/10.1016/j.ssc.2007.03.012.

[76] K. Puzniak, C. Aguilar-Maldonado, R. Feyerherm, K. Prokeš, A.T.M.N. Islam, Y. Skourski, L. Keller, B. Lake, Magnetic structure and phase diagram of the Heisenberg-Ising spin chain antiferromagnetic $PbCo_2V_2O_8$, Phys Rev B 108 (2023) 144432. https://doi.org/10.1103/PhysRevB.108.144432.

[77] H. Fjellvå, E. Gulbrandsen, S. Aasland, A. Olsen, B.C. Hauback, Crystal Structure and Possible Charge Ordering in One-Dimensional $Ca_3Co_2O_6$, J Solid State Chem 124 (1996) 190–194. https://doi.org/10.1006/jssc.1996.0224.

[78] C. Mazzoli, A. Bombardi, S. Agrestini, M.R. Lees, Resonant X-ray scattering study of $Ca_3Co_2O_6$ ground state: Preliminary results of magnetic field effects, Physica B Condens Matter 404 (2009) 3042–3044. https://doi.org/10.1016/j.physb.2009.07.008.

[79] S. Aasland, H. Fjellvåg, B. Hauback b, Magnetic properties of the one-dimensional $Ca_3Co_2O_6$, Solid State Commun 101 (1997) 187–192. https://doi.org/10.1016/S0038-1098(96)00531-5.

[80] H. Kageyama, K. Yoshimura, K. Kosuge, M. Azuma, M. Takano, H. Mitamura, T. Goto, Magnetic Anisotropy of $Ca_3Co_2O_6$ with Ferromagnetic Ising Chains, J Physical Soc Japan 66 (1997) 3996–4000. https://doi.org/10.1143/JPSJ.66.3996.

[81] A. Maignan, C. Michel, A.C. Masset, C. Martin, B. Raveau, Single crystal study of the one dimensional $Ca_3Co_2O_6$ compound: five stable configurations for the Ising triangular lattice, Eur. Phys. J. B 15 (2000) 657–663. https://doi.org/10.1007/PL00011051.

[82] A. Maignan, V. Hardy, S. Hébert, M. Drillon, M.R. Lees, O. Petrenko, D.M.K. Paul, D. Khomskii, Quantum tunneling of the magnetization in the Ising chain compound $Ca_3Co_2O_6$, J Mater Chem 14 (2004) 1231–1234. https://doi.org/10.1039/b316717h.

[83] R. Vidya, P. Ravindran, H. Fjellvåg, A. Kjekshus, O. Eriksson, Tailor-made electronic and magnetic properties in one-dimensional pure and Y-substituted $Ca_3Co_2O_6$, Phys Rev Lett 91 (2003) 186404. https://doi.org/10.1103/PhysRevLett.91.186404.

[84] V. Hardy, S. Lambert, R. Lees, D. McK. Paul, Specific heat and magnetization study on single crystals of the frustrated quasi-one-dimensional oxide $Ca_3Co_2O_6$, Phys Rev B 68 (2003) 014424. https://doi.org/10.1103/PhysRevB.68.014424.

[85] V. Eyert, C. Laschinger, T. Kopp, R. Frésard, Extended moment formation and magnetic ordering in the trigonal chain compound $Ca_3Co_2O_6$, Chem Phys Lett 385 (2004) 249–254. https://doi.org/10.1016/j.cplett.2003.12.105.

[86] H. Wu, M.W. Haverkort, Z. Hu, D.I. Khomskii, L.H. Tjeng, Nature of magnetism in $Ca_3Co_2O_6$, Phys Rev Lett 95 (2005) 186401. https://doi.org/10.1103/PhysRevLett.95.186401.

[87] K. Takubo, T. Mizokawa, S. Hirata, J.Y. Son, A. Fujimori, D. Topwal, D.D. Sarma, S. Rayaprol, E. V. Sampathkumaran, Electronic structure of $Ca_3CoXO_6$ (X=Co, Rh, Ir) studied by x-ray photoemission spectroscopy, Phys Rev B 71 (2005) 073406. https://doi.org/10.1103/PhysRevB.71.073406.

[88] T. Burnus, Z. Hu, M.W. Haverkort, J.C. Cezar, D. Flahaut, V. Hardy, A. Maignan, N.B. Brookes, A. Tanaka, H.H. Hsieh, H.J. Lin, C.T. Chen, L.H. Tjeng, Valence, spin, and orbital state of Co ions in one-dimensional $Ca_3Co_2O_6$: An X-ray absorption and magnetic circular dichroism study, Phys Rev B 74 (2006) 245111. https://doi.org/10.1103/PhysRevB.74.245111.

[89] L.C. Chapon, Origin of the long-wavelength magnetic modulation in $Ca_3Co_2O_6$, Phys Rev B 80 (2009) 172405. https://doi.org/10.1103/PhysRevB.80.172405.

[90] T. Moyoshi, K. Motoya, Incommensurate magnetic structure and its long-time variation in a geometrically frustrated magnet $Ca_3Co_2O_6$, J Physical Soc Japan 80 (2011) 034701. https://doi.org/10.1143/JPSJ.80.034701.

[91] S. Agrestini, C. Mazzoli, A. Bombardi, M.R. Lees, Incommensurate magnetic ground state revealed by resonant x-ray scattering in the frustrated spin system $Ca_3Co_2O_6$, Phys Rev B 77 (2008) 140403(R). https://doi.org/10.1103/PhysRevB.77.140403.

[92] P. Lampen, N.S. Bingham, M.H. Phan, H. Srikanth, H.T. Yi, S.W. Cheong, Macroscopic phase diagram and magnetocaloric study of metamagnetic transitions in the spin chain system $Ca_3Co_2O_6$, Phys Rev B 89 (2014) 144414. https://doi.org/10.1103/PhysRevB.89.144414.

[93] H. Kageyama, K. Yoshimura, K. Kosuge, H. Mitamura, T. Goto, Field-Induced Magnetic Transitions in the One-Dimensional Compound $Ca_3Co_2O_6$, J Physical Soc Japan 66 (1997) 1607–1610. https://doi.org/10.1143/JPSJ.66.1607.

[94] V. Hardy, M.R. Lees, O.A. Petrenko, D.M.K. Paul, D. Flahaut, S. Hébert, A. Maignan, Temperature and time dependence of the field-driven magnetization steps in $Ca_3Co_2O_6$ single crystals, Phys Rev B 70 (2004) 064424. https://doi.org/10.1103/PhysRevB.70.064424.

[95] Y.B. Kudasov, Steplike magnetization in a spin-chain system: $Ca_3Co_2O_6$, Phys Rev Lett 96 (2006) 027212. https://doi.org/10.1103/PhysRevLett.96.027212.

[96] C.L. Fleck, M.R. Lees, S. Agrestini, G.J. McIntyre, O.A. Petrenko, Field-driven magnetisation steps in $Ca_3Co_2O_6$ : A single-crystal neutron-diffraction study, Europhys Lett 90 (2010) 67006. https://doi.org/10.1209/0295-5075/90/67006.

[97] X.Y. Yao, S. Dong, J.M. Liu, Steplike magnetization of spin chains in a triangular lattice: Monte Carlo simulations, Phys Rev B 73 (2006) 212415. https://doi.org/10.1103/PhysRevB.73.212415.

[98] X. Yao, S. Dong, H. Yu, J. Liu, Monte Carlo simulation of magnetic behavior of a spin-chain system on a triangular lattice, Phys Rev B 74 (2006) 134421. https://doi.org/10.1103/PhysRevB.74.134421.

[99] M.H. Qin, K.F. Wang, J.M. Liu, Two-step magnetization in a spin-chain system on the triangular lattice: Wang-Landau simulation, Phys Rev B 79 (2009) 172405. https://doi.org/10.1103/PhysRevB.79.172405.

[100] R. Soto, G. Martínez, M.N. Baibich, J.M. Florez, P. Vargas, Metastable states in the triangular-lattice Ising model studied by Monte Carlo simulations: Application to the spin-chain compound $Ca_3Co_2O_6$, Phys Rev B 79 (2009) 184422. https://doi.org/10.1103/PhysRevB.79.184422.

[101] Y.B. Kudasov, A.S. Korshunov, V.N. Pavlov, D.A. Maslov, Dynamics of magnetization in frustrated spin-chain system $Ca_3Co_2O_6$, Phys Rev B 78 (2008) 132407. https://doi.org/10.1103/PhysRevB.78.132407.

[102] J.A.M. Paddison, S. Agrestini, M.R. Lees, C.L. Fleck, P.P. Deen, A.L. Goodwin, J.R. Stewart, O.A. Petrenko, Spin correlations in $Ca_3Co_2O_6$: Polarized-neutron diffraction and Monte Carlo study, Phys Rev B 90 (2014) 014411. https://doi.org/10.1103/PhysRevB.90.014411.

[103] I. Nekrashevich, X. Ding, F. Balakirev, H.T. Yi, S.W. Cheong, L. Civale, Y. Kamiya, V.S. Zapf, Reaching the equilibrium state of the frustrated triangular Ising magnet $Ca_3Co_2O_6$, Phys Rev B 105 (2022) 024426. https://doi.org/10.1103/PhysRevB.105.024426.

[104] Y. Kamiya, C.D. Batista, Formation of magnetic microphases in $Ca_3Co_2O_6$, Phys Rev Lett 109 (2012) 067204. https://doi.org/10.1103/PhysRevLett.109.067204.

[105] Y. Kamiya, Magnetic field induced deformation of the spin density wave microphases in $Ca_3Co_2O_6$, Phys Rev B 107 (2023) 134409. https://doi.org/10.1103/PhysRevB.107.134409.

[106] S. Agrestini, C.L. Fleck, L.C. Chapon, C. Mazzoli, A. Bombardi, M.R. Lees, O.A. Petrenko, Slow magnetic order-order transition in the spin chain antiferromagnet $Ca_3Co_2O_6$, Phys Rev Lett 106 (2011) 197204. https://doi.org/10.1103/PhysRevLett.106.197204.

[107] K. Motoya, T. Kihara, H. Nojiri, Y. Uwatoko, M. Matsuda, T. Hong, Time and magnetic field variations of magnetic structure in the triangular lattice magnet $Ca_3Co_2O_6$, J Physical Soc Japan 87 (2018). https://doi.org/10.7566/JPSJ.87.114703.

[108] D.P. Kozlenko, N.T. Dang, N.O. Golosova, S.E. Kichanov, E. V. Lukin, P.J. Lampen Kelley, E.M. Clements, K. V. Glazyrin, S.H. Jabarov, T.L. Phan, B.N. Savenko, H. Srikanth, M.H. Phan, Pressure-induced modifications of the magnetic order in the spin-chain compound $Ca_3Co_2O_6$, Phys Rev B 94 (2018) 134435. https://doi.org/10.1103/PhysRevB.98.134435.

[109] B. Leedahl, M. Sundermann, A. Amorese, A. Severing, H. Gretarsson, L. Zhang, A.C. Komarek, A. Maignan, M.W. Haverkort, L.H. Tjeng, Origin of Ising magnetism in $Ca_3Co_2O_6$ unveiled by orbital imaging, Nat Commun 10 (2019) 5447. https://doi.org/10.1038/s41467-019-13273-4.

[110] G. Allodi, P. Santini, S. Carretta, S. Agrestini, C. Mazzoli, A. Bombardi, M.R. Lees, R. De Renzi, Exchange interactions in $Ca_3Co_2O_6$ probed locally by NMR, Phys Rev B 89 (2014) 104401. https://doi.org/10.1103/PhysRevB.89.104401.

[111] J. Sugiyama, H. Nozaki, Y. Ikedo, K. Mukai, D. Andreica, A. Amato, J.H. Brewer, E.J. Ansaldo, G.D. Morris, T. Takami, H. Ikuta, Evidence of two dimensionality in quasi-one-dimensional cobalt oxides, Phys Rev Lett 96 (2006) 197206. https://doi.org/10.1103/PhysRevLett.96.197206.

[112] Y.J. Choi, H.T. Yi, S. Lee, Q. Huang, V. Kiryukhin, S.W. Cheong, Ferroelectricity in an ising chain magnet, Phys Rev Lett 100 (2008) 047601. https://doi.org/10.1103/PhysRevLett.100.047601.

[113] H. Wu, T. Burnus, Z. Hu, C. Martin, A. Maignan, J.C. Cezar, A. Tanaka, N.B. Brookes, D.I. Khomskii, L.H. Tjeng, Ising magnetism and ferroelectricity in $Ca_3CoMnO_6$, Phys Rev Lett 102 (2009) 026404. https://doi.org/10.1103/PhysRevLett.102.026404.

[114] J.W. Kim, E.D. Mun, X. Ding, A. Hansen, M. Jaime, N. Harrison, H.T. Yi, Y. Chai, Y. Sun, S.W. Cheong, V.S. Zapf, Metastable states in the frustrated triangular compounds $Ca_3Co_{2-x}Mn_xO_6$ and $Ca_3Co_2O_6$, Phys Rev B 98 (2018) 024407. https://doi.org/10.1103/PhysRevB.98.024407.

[115] S. Niitaka, K. Yoshimura, K. Kosuge, M. Nishi, K. Kakurai, Partially disordered antiferromagnetic phase in $Ca_3CoRhO_6$, Phys Rev Lett 87 (2001) 177202. https://doi.org/10.1103/PhysRevLett.87.177202.

[116] M.-H. Whangbo, D. Dai, H.-J. Koo, S. Jobic, Investigations of the oxidation states and spin distributions in $Ca_3Co_2O_6$ and $Ca_3CoRhO_6$ by spin-polarized electronic band structure calculations, Solid State Commun 125 (2003) 413–417. https://doi.org/10.1016/S0038-1098(02)00872-4.

[117] J. An, C.W. Nan, Electronic structure and transport of $Ca_3Co_2O_6$ and $Ca_3CoNiO_6$, Solid State Commun 129 (2004) 51–56. https://doi.org/10.1016/j.ssc.2003.09.013.

[118] T. Takami, H. Ikuta, U. Mizutani, Thermoelectric properties of $A_{n+2}Co_{n+1}O_{3n+3}$ (A = Ca, Sr, Ba, $n$ = 1-5), Jpn J Appl Phys 43 (2004) 8208–8212. https://doi.org/10.1143/JJAP.43.8208.

[119] J. Sugiyama, H. Nozaki, J.H. Brewer, E.J. Ansaldo, T. Takami, H. Ikuta, U. Mizutani, Appearance of a two-dimensional antiferromagnetic order in quasi-one-dimensional cobalt oxides, Phys Rev B 72 (2005) 064418. https://doi.org/10.1103/PhysRevB.72.064418.

[120] J. Sugiyama, H. Nozaki, Y. Ikedo, K. Mukai, D. Andreica, A. Amato, J.H. Brewer, E.J. Ansaldo, G.D. Morris, T. Takami, H. Ikuta, Two dimensionality in quasi-one-dimensional cobalt oxides

confirmed by muon-spin spectroscopy, J Magn Magn Mater 310 (2007) 2719–2721. https://doi.org/10.1016/j.jmmm.2006.10.1142.

[121] M.A. Melkozerova, G. V. Bazuev, Quasi-one-dimensional oxides $Sr_4Co_{3-x}Mn_xO_9$ ($0 \leq x \leq 3$): Synthesis and magnetic properties, Russian Journal of Inorganic Chemistry 51 (2006) 362–367. https://doi.org/10.1134/S0036023606030053.

[122] K. Boulahya, M. Parras, J.M. González-Calbet, J.L. Martínez, Synthesis, structural characterization, and magnetic study of $Sr_4Mn_2CoO_9$, Chemistry of Materials 15 (2003) 3537–3542. https://doi.org/10.1021/cm034130t.

[123] K. Boulahya, M. Hernando, M. Parras, J.M. González-Calbet, New stabilized phases in the Sr/Ca-Mn-Co-O system: Structural-magnetic properties relationship, J Mater Chem 17 (2007) 1620–1626. https://doi.org/10.1039/b613844f.

[124] M.M. Seikh, V. Caignaert, O. Perez, B. Raveau, V. Hardy, Single-ion and single-chain magnetism in triangular spin-chain oxides, Phys Rev B 95 (2017) 174417. https://doi.org/10.1103/PhysRevB.95.174417.

[125] M. Hernando, K. Boulahya, M. Parras, A. Varela, J.M. González-Calbet, J.L. Martínez, Structural and magnetic study of $Sr_{3.3}Ca_{0.7}CoRh_2O_9$: A new partially ordered antiferromagnetic system, Chemistry of Materials 14 (2002) 4948–4954. https://doi.org/10.1021/cm0211756.

[126] T.I. Chupakhina, M.A. Melkozerova, G. V. Bazuev, Phase formation features and magnetic properties of complex oxides in the systems Sr-Co-M-O (M = Zn, Cu), Russian Journal of Inorganic Chemistry 58 (2013) 253–258. https://doi.org/10.1134/S0036023613030054.

[127] A.J. Neer, J. Milam-Guerrero, J.E. So, B.C. Melot, K.A. Ross, Z. Hulvey, C.M. Brown, A.A. Sokol, D.O. Scanlon, Ising-like antiferromagnetism on the octahedral sublattice of a cobalt-containing garnet and the potential for quantum criticality, Phys Rev B 95 (2017) 144419. https://doi.org/10.1103/PhysRevB.95.144419.

[128] A.J. Neer, J.A. Milam-Guerrero, V.A. Fischer, M. Zheng, N.R. Spence, C. Cozzan, M. Gu, J.M. Rondinelli, C.M. Brown, B.C. Melot, Magnetic-Field-Induced Dielectric Anomalies in Cobalt-Containing Garnets, Inorg Chem 61 (2022) 5452–5458. https://doi.org/10.1021/acs.inorgchem.1c03792.

[129] I.W. Johnstone, D.J. Lockwood, M.W.C. Dharma-wardana, Influence of interchain coupling on the one-dimensional magnon Raman spectrum of $CsCoBr_3$, Solid State Commun 36 (1980) 593–597. https://doi.org/10.1016/0038-1098(80)90094-0.

[130] W.P. Lehmann, W. Breitlingc, R. Weber, Raman scattering study of spin dynamics in the quasi-1D Ising antiferromagnets $CsCoCl_3$, and $CsCoBr_3$, J. Phys. C: Solid State Phys 14 (1981) 4655–4676. https://doi.org/10.1088/0022-3719/14/31/014.

[131] D.J. Lockwood, I.W. Johnstone, Raman scattering from magnons and excitons in the 3-D ordered phases of CsCoBr$_3$, J Appl Phys 53 (1982) 8169–8171. https://doi.org/10.1063/1.330281.

[132] F. Matsubara, S. Inawashiro, H. Ohhara, On the magnetic Raman scattering in CsCoCl3, CsCoBr$_3$ and RbCoCl$_3$, J. Phys.: Condens. Matter 3 (1991) 1815–1826. https://doi.org/10.1088/0953-8984/3/12/012.

[133] S.E. Nagler, W.J.L. Buyers, R.L. Armstrong, B. Briat, Ising-like spin-, quasi-one-dimensional antiferromagnets: Spin-wave response in CsCoX$_3$ salts, Phys Rev B 27 (1983) 1784–1799. https://doi.org/10.1103/PhysRevB.27.1784.

[134] S.E. Nagler, W.J.L. Buyers, R.L. Armstrong, B. Briat, Propagating Domain Walls in CsCoBr$_3$, Phys Rev Lett 49 (1982) 590–592.

[135] H. Yoshizawa', K. Hirakawa, S.K. Satija, G. Shirane, Dynamical correlation functions in a one-dimensional Ising-like antiferromagnetic CsCoCl$_3$. A neutron scattering study, Phys Rev B 23 (1981) 2298–2307. https://doi.org/10.1103/PhysRevB.23.2298.

[136] J. Villain, Propagative spin relaxation in the Ising-like antiferromagnetic linear chain, Physica B+C 79 (1975) 1–12. https://doi.org/10.1016/0378-4363(75)90101-1.

[137] S.E. Nagler, W.J.L. Buyers, R.L. Armstrong, B. Briat, Solitons in the one-dimensional antiferromagnet CsCoBr$_3$, Phys Rev B 28 (1983) 3873–3885. https://doi.org/10.1103/PhysRevB.28.3873.

[138] H. Shiba, Quantization of Magnetic Excitation Continuum Due to lnterchain Coupling in Nearly One-Dimensional Ising-Like Antiferromagnets, Progress of Theoretical Physics 64 (1980) 466–478. https://doi.org/10.1143/PTP.64.466.

[139] N. Ishimura, H. Shiba, Dynamical Correlation Functions of One-Dimensional Anisotropic Heisenberg Model with Spin 1/2. I: Ising-Like Antiferromagnets, Progress of Theoretical Physics 63 (1980) 743–758. https://doi.org/10.1143/PTP.63.743.

[140] F. Matsubara, S. Inawashiro, Pair States and Bound States of Solitons in an Ising-Like S=1/2 Antiferromagnet with a Weak Next-Nearest-Neighbor Interaction on a Linear Chain, J Physical Soc Japan 58 (1989) 4284–4287. https://doi.org/10.1143/JPSJ.58.4284.

[141] J.P. Goff, D.A. Tennant, S.E. Nagler, Exchange mixing and soliton dynamics in the quantum spin chain CsCoCl$_3$, Phys Rev B 52 (1995) 15992–16000. https://doi.org/10.1103/PhysRevB.52.15992.

[142] R. Jorke, U. Durr, Properties of magnetic excitations in RbCoCl$_3$, J. Phys. C: Solid State Phys. 16 (1983) L1129–L1136. https://doi.org/10.1088/0022-3719/16/31/007.

[143] D.J. Lockwood, I.W. Johnstone, H.J. Labbet, B. Briat, Raman scattering from the 1D antiferromagnets $RbCoCl_3$ and $RbNiCl_3$, J. Phys. C: Solid State Phys. 16 (1983) 6451–6474. https://doi.org/10.1088/0022-3719/16/33/018.

[144] M. Mena, N. Hänni, S. Ward, E. Hirtenlechner, R. Bewley, C. Hubig, U. Schollwöck, B. Normand, K.W. Krämer, D.F. McMorrow, C. Rüegg, Thermal Control of Spin Excitations in the Coupled Ising-Chain Material $RbCoCl_3$, Phys Rev Lett 124 (2020) 257201. https://doi.org/10.1103/PhysRevLett.124.257201.

[145] N.P. Hänni, D. Sheptyakov, M. Mena, E. Hirtenlechner, L. Keller, U. Stuhr, L.P. Regnault, M. Medarde, A. Cervellino, C. Rüegg, B. Normand, K.W. Krämer, Magnetic order in the quasi-one-dimensional Ising system $RbCoCl_3$, Phys Rev B 103 (2021) 094424. https://doi.org/10.1103/PhysRevB.103.094424.

[146] M.G. Cottam, D.J. Lockwood, Zeeman-ladder analysis of the Raman magnon energies in the quasi-one-dimensional antiferromagnet $RbCoCl_3$, Phys Rev B 105 (2022) 064411. https://doi.org/10.1103/PhysRevB.105.064411.

[147] M.G. Cottam, D.J. Lockwood, Unusual behaviour of the spin-phonon coupling in the quasi-one-dimensional antiferromagnet $RbCoCl_3$, Sci Rep 12 (2022) 14065. https://doi.org/10.1038/s41598-022-18073-3.

[148] S. Calder, L.D. Sanjeewa, V.O. Garlea, J. Xing, R.J. Terry, J.W. Kolis, Magnetic interactions in the one-dimensional spin-chain metal-organic compounds $M(N_2H_5)_2(SO_4)_2$ (M=Cu, Co, Mn), Phys Rev Mater 6 (2022) 124407. https://doi.org/10.1103/PhysRevMaterials.6.124407.

[149] M. Böhme, M. Rams, C. Krebs, S. Mangelsen, I. Jess, W. Plass, C. Näther, $Co(NCS)_2$ Chain Compound with Alternating 5- and 6-Fold Coordination: Influence of Metal Coordination on the Magnetic Properties, Inorg Chem 61 (2022) 16841–16855. https://doi.org/10.1021/acs.inorgchem.2c02813.